\documentclass[11pt]{article}
\usepackage{amssymb,amsmath,amsfonts}
\usepackage{graphicx}
\usepackage{graphics}
\usepackage{eepic,epsfig}

\@addtoreset{equation}{section}

\makeatother

\begin{document}

\title{The Casimir effect for fermionic currents in conical rings with
applications to graphene ribbons}
\author{S. Bellucci$^{1}$, I. Brevik$^{2}$, A. A. Saharian$^{3}$, H. G.
Sargsyan$^{3}$ \vspace{0.3cm} \\
\textit{$^1$ INFN, Laboratori Nazionali di Frascati,}\\
\textit{Via Enrico Fermi 40, 00044 Frascati, Italy} \vspace{0.3cm}\\
\textit{$^{2}$Department of Energy and Process Engineering,}\\
\textit{Norwegian University of Science and Technology, Trondheim, Norway}
\vspace{0.3cm}\\
\textit{$^{3}$Department of Physics, Yerevan State University,}\\
\textit{1 Alex Manoogian Street, 0025 Yerevan, Armenia}}
\maketitle

\begin{abstract}
We investigate the combined effects of boundaries and topology on the vacuum
expectation values (VEVs) of the charge and current densities for a massive
2D fermionic field confined on a conical ring threaded by a magnetic flux.
Different types of boundary conditions on the ring edges are considered for
fields realizing two inequivalent irreducible representations of the
Clifford algebra. The related bound states and zero energy fermionic modes
are discussed. The edge contributions to the VEVs of the charge and
azimuthal current densities are explicitly extracted and their behavior in
various asymptotic limits is considered. On the ring edges the azimuthal
current density is equal to the charge density or has an opposite sign. We
show that the absolute values of the charge and current densities increase
with increasing planar angle deficit. Depending on the boundary conditions,
the VEVs are continuous or discontinuous at half-integer values of the ratio
of the effective magnetic flux to the flux quantum. The discontinuity is
related to the presence of the zero energy mode. By combining the results
for the fields realizing the irreducible representations of the Clifford
algebra, the charge and current densities are studied in parity and
time-reversal symmetric fermionic models. If the boundary conditions and the
phases in quasiperiodicity conditions for separate fields are the same the
total charge density vanishes. Applications are given to graphitic cones
with edges (conical ribbons).
\end{abstract}

\section{Introduction}

In the last decade the two-dimensional (2D) fermionic models have attracted
considerable attention, both from the experimental and theoretical points of
view. Besides being simplified models in particles physics, they also appear
as effective theories describing low-energy excitations of the electronic
subsystem in a number of condensed matter systems \cite{Naga99}-\cite{Mari17}%
. The condensed matter realizations of 2D fermions include Weyl semimetals,
graphene family materials (graphene, silicene, germanene, stanene),
topological insulators, high-temperature superconductors and d-density-wave
states. The dynamics of the low-energy charge carriers in these systems is
governed by the Dirac equation with the Fermi velocity appearing instead of
the velocity of light \cite{Gusy07}-\cite{Xiao11}. Other examples of the
systems with Dirac fermions include ultracold atoms confined by lattice
potentials, nano-patterned 2D electron gases and photonic crystals. An
important advantage with these artificial systems is that the corresponding
symmetry and parameters are relatively easy to control. This provides new
opportunities for studying the influence of those parameters on the dynamics
of Dirac quasiparticles. The interesting effects induced by the change of
the parameters include topological phase transitions, merging of the Dirac
points, generation of the anisotropy of the hopping parameters.

The emergence of Dirac fermions in condensed matter systems provides an
interesting possibility to observe different kinds of effects in the system
of interacting fields. Here we have a situation typical for braneworld
models in high-energy physics where a part of the fields are confined on
hypersurface (branes) whereas other fields propagate in the bulk. An example
is the set of 2D fermionic and 3D electromagnetic field. In quantum field
theory, the interaction of the fermionic field, confined on a surface, with
the fluctuations of the bulk quantized fields gives rise to the Casimir type
shifts in the expectation values of physical observables (for the Casimir
effect and its applications in high-energy and condensed matter physics see
\cite{Eliz94}). In recent years, the Casimir effect in systems involving
graphene structures as boundaries have seen novel developments (see \cite%
{Bord09,Rodr17} and \cite{Klim09} for reviews). In \cite{Rodr17} it has been
shown that the various electronic phases of graphene family materials,
tunable by external fields, lead to different scaling laws and significant
magnitude changes for the Casimir forces. These features can be used to
probe the 2D Dirac physics of the corresponding materials. The topologically
and boundary induced effects in interacting fermionic systems were discussed
in \cite{Eliz94b,Flac12}.

In Refs. \cite{Bord09,Rodr17,Klim09} the Casimir effect is considered for
the electromagnetic field. The role of the 2D fermionic field was reduced to
the generation of boundary condition on the quantized electromagnetic field.
In graphene family materials with edges (nanoribbons) or with nontrivial
spatial topology (nanotubes and nanorings) the Casimir type effects appear
for the quantum 2D fermionic field as well. The topological Casimir effect
for the fermionic condensate, for the vacuum expectation values (VEVs) of
the energy-momentum tensor and of the current density in cylindrical and
toroidal nanotubes has been investigated in \cite{Bell09,Bell10}. The finite
temperature effects were discussed in \cite{Bell14}. In finite length
nanotubes, in addition to the topological parts, edge-induced Casimir
contributions are present. In carbon nanotubes, these contributions depend
on the chirality of the tube and have been studied in \cite{Bell09b}-\cite%
{Bell13}. The Casimir effect in a more complicated geometry of hemisphere
capped tubes was considered in \cite{Beze16}. The condensed matter
realizations of 2D fermions with curved geometries can be used to model the
influence of the gravitational field on the quantum matter (for various
types of mechanisms of the generation of curvature in graphene and the
related effects see \cite{Kole09}). Both the topological and
boundary-induced Casimir effects for the charge and current densities of a
fermionic field confined on curved graphene tubes with locally anti-de
Sitter geometry have been discussed in \cite{Bell17}.

In the present paper we investigate the effects of planar angle deficit on
the VEVs of the charge and current densities for a 2D fermionic field
confined on a conical ring threaded by a magnetic flux. Among the condensed
matter realizations of this system are the graphitic cones. These structures
are obtained from a graphene sheet by cutting one or more sectors with the
angle $\pi /3$ and gluing the two edges of the remaining sector. The
corresponding planar angle deficit is given by $\pi n_{c}/3$, with $%
n_{c}=1,2,\ldots ,5$ being the number of the removed sectors. The graphitic
cones with all these values of the angle deficit were observed
experimentally in both the forms as caps on the ends of the nanotubes and as
free-standing structures (see, for instance, \cite{Kris97}). The electronic
properties of graphitic cones have been discussed in \cite{Lamm00}-\cite%
{Chak11}. The background geometry under consideration in the present paper
with 2D fermionic field corresponds to the continuum description of finite
radius graphitic cones with cutted apex. Some limiting cases have been
considered previously in the literature. The vacuum polarization effects in
the boundary-free geometry with applications to graphitic cones have been
discussed in \cite{Site08b}-\cite{Beze12}. The zero temperature fermionic
condensate, the expectation values of the charge and current densities and
of the energy-momentum tensor for a conical geometry with a single circular
boundary where studied in \cite{Bell11}-\cite{Beze12}. The combined effects
of the edge and of finite temperature have been considered in \cite{Bell16T}%
. The ground state fermionic charge and current densities in planar rings
were investigated in \cite{Bell16}.

The organization of the paper is as follows. In the next section the field,
background geometry and the mode functions for a fermionic field are
presented. In section \ref{sec:jmu} these modes are used for the evaluation
of the VEVs of the charge and current densities in conical rings. Different
representations of the VEVs are given and their properties are investigated.
Several limiting cases and asymptotics are discussed in section \ref%
{sec:Asymp}. Numerical examples for the behavior of both the charge and
current densities are presented. The charge and current densities for the
fermionic field realizing the second irreducible representation of the
Clifford algebra are considered in section \ref{sec:Tsym}. Applications are
given to 2D fermionic systems with parity and time-reversal symmetry and to
graphene nanocones. The main results are summarized in section \ref{sec:Conc}%
. The bound states for different boundary conditions on the edges of the
ring and their contributions to the VEVs of the charge and current densities
are discussed in appendix \ref{sec:App1}. In appendix \ref{sec:Sp} we
consider the contribution of the special mode for half-integer values of the
parameter related to the enclosed magnetic flux and to the phase in the
periodicity condition along the azimuthal direction.

\section{Problem setup and the fermionic modes}

\label{sec:modes}

For the background geometry under consideration the (2+1)-dimensional line
element is given by
\begin{equation}
ds^{2}=g_{\mu \nu }dx^{\mu }dx^{\nu }=dt^{2}-dr^{2}-r^{2}d\phi ^{2}\ ,
\label{linel}
\end{equation}%
where the cylindrical spatial coordinates $r$ and $\phi $ vary in the ranges
$r\geqslant 0$ and $0\leqslant \phi \leqslant \phi _{0}$. The special case $%
\phi _{0}=2\pi $ corresponds to the (2+1)-dimensional Minkwoski spacetime
described in cylindrical coordinates. For $\phi _{0}<2\pi $, the line
element describes a cone with planar angle deficit $2\pi -\phi _{0}$ and
with the apex at $r=0$.

As a quantum field we consider a charged fermionic field $\psi (x)$ in the
irreducible representation of the Clifford algebra. The latter is realized
by two-component spinors. Additionally, the presence of an external
classical abelian gauge field $A_{\mu }$ will be assumed. The dynamics of
the field is governed by the Dirac equation
\begin{equation}
\left( i\gamma ^{\mu }D_{\mu }-sm\right) \psi (x)=0.  \label{Dirac}
\end{equation}%
The gauge extended covariant derivative is defined as $D_{\mu }=\partial
_{\mu }+\Gamma _{\mu }+ieA_{\mu }$, with $\Gamma _{\mu }$ being the spin
connection and $e$ being the charge of the field quanta. In (\ref{Dirac}) we
have introduced the parameter $s$, with the values $s=+1$ and $s=-1$,
corresponding to two inequivalent irreducible representations of the
Clifford algebra in $(2+1)$-dimensions (see also section \ref{sec:Tsym}). In
the coordinate system under consideration for the Dirac matrices in (\ref%
{Dirac}) we use the representation
\begin{equation}
\gamma ^{0}=\left(
\begin{array}{cc}
1 & 0 \\
0 & -1%
\end{array}%
\right) ,\quad \gamma ^{l}=\frac{i^{2-l}}{r^{l-1}}\left(
\begin{array}{cc}
0 & e^{-iq\phi } \\
(-1)^{l-1}e^{iq\phi } & 0%
\end{array}%
\right) ,  \label{gamma}
\end{equation}%
where $l=1,2$ and $q=2\pi /\phi _{0}$.

\begin{figure}[tbph]
\begin{center}
\epsfig{figure=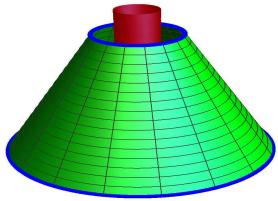,width=4.5cm,height=4cm}
\end{center}
\caption{Conical ring with the edges $r=a$ and $r=b$ threaded by a magnetic
flux.}
\label{fig1}
\end{figure}

It will be assumed that the field is confined in the region $a\leq r\leq b$
(conical ring, the geometry of the problem is depicted in figure \ref{fig1}%
). On the edges of the ring the boundary conditions
\begin{equation}
\left( 1+i\lambda _{r}n_{\mu }\gamma ^{\mu }\right) \psi (x)=0,\;r=a,b,
\label{BCs}
\end{equation}%
will be imposed. Here $n_{\mu }$ is the inward pointing unit vector normal
to the boundary and the parameters $\lambda _{a}$ and $\lambda _{b}$ take
the values $\pm 1$. For the boundary at $r=u$, $u=a,b$, and in the region
under consideration the normal is given by $n_{\mu }=n_{u}\delta _{\mu }^{1}$%
, where%
\begin{equation}
n_{a}=-1,\;n_{b}=1.  \label{nab}
\end{equation}%
It can be shown that, as a consequence of the conditions (\ref{BCs}), on the
boundaries we get $n_{\mu }j^{\mu }=0$ with $j^{\mu }=e\bar{\psi}\gamma
^{\mu }\psi $ being the current density and $\bar{\psi}=\psi ^{\dagger
}\gamma ^{0}$ is the Dirac adjoint. This means that the normal component of
the fermionic current vanishes on the edges and, consequently, the dynamics
is completely determined by the field equation and the boundary conditions.
The special case with $\lambda _{r}=1$, $r=a,b$, corresponds to the MIT bag
boundary condition (or infinite mass boundary condition in the condensed
matter context) on both the edges. Comparing the analytical results on the
electronic properties of circular graphene quantum dots derived within the
Dirac model with the bag boundary condition to those obtained from the
tight-binding model, the authors of \cite{Schn08} have found a good
qualitative agreement between those two approaches. Considering different
boundary conditions in the continuous model for graphene devices and
comparing with the experiments, a similar conclusion is made in \cite{Bene12}%
. Another special case with $\lambda _{r}=-1$ was considered in \cite{Berr87}%
. More general boundary conditions for the confinement of fermions and their
realizations in graphene made structures have been discussed in \cite{Mcca04}%
.

The background geometry has nontrivial topology and, in addition to the
boundary conditions on the ring edges, one needs to specify the periodicity
condition along the azimuthal direction. We will assume the condition
\begin{equation}
\psi (t,r,\phi +\phi _{0})=e^{2\pi i\chi }\psi (t,r,\phi ),  \label{PC}
\end{equation}%
with a general phase $2\pi \chi $. The special cases $\chi =0$ and $\chi
=1/2 $ correspond to untwisted and twisted fermionic fields. The values for
the parameter $\chi $ realized in graphene cones will be discussed in
section \ref{sec:Tsym}. As it will be seen below, the nontrivial phase in (%
\ref{PC}) can be interpreted in terms of the fictitious flux threading the
ring.

Here we are interested in the VEVs\ of the charge and current densities
induced by a magnetic flux threading the conical ring. The magnetic field is
localized inside the region $r<a$ and its influence on the characteristics
of the fermionic vacuum is purely topological. This is an Aharonov-Bohm type
effect related to the nontrivial topology of the background space. In the
region under consideration, $a\leq r\leq b$, the covariant components of the
vector potential of the gauge field in the system of coordinates $(t,r,\phi
) $ are given by $A_{\mu }=(0,0,A)$. Note that for the corresponding
physical component one has $A_{\phi }=-A/r$. The magnetic flux enclosed by
the ring is expressed in terms of the covariant component as $\Phi =-\phi
_{0}A$. The physical effects on the ring are completely determined by this
flux and they do not depend on the radial distribution of the flux in the
region $r<a$.

The VEV of the current density, $\langle 0|j^{\mu }(x)|0\rangle \equiv
\langle j^{\mu }(x)\rangle $, can be evaluated by using the relation%
\begin{equation}
\langle j^{\mu }(x)\rangle =-\frac{e}{2}\mathrm{Tr}(\gamma ^{\mu
}S^{(1)}(x,x)),  \label{VEVj}
\end{equation}%
where the trace in the right-hand side is over spinor indices and $%
S^{(1)}(x,x^{\prime })$ is the fermion two-point function. Its spinorial
components, with spinor indices $i$ and $k$, are defined as the VEV $%
S_{ik}^{(1)}(x,x^{\prime })=\langle 0|[\psi _{i}(x),\bar{\psi}_{k}(x^{\prime
})]|0\rangle $. Let $\{\psi _{\sigma }^{(+)}(x),\psi _{\sigma }^{(-)}(x)\}$
be the complete set of the positive and negative energy fermionic mode
functions, obeying the field equation (\ref{Dirac}), the boundary conditions
(\ref{BCs}) and the periodicity condition (\ref{PC}). They are specified by
the set of quantum numbers $\sigma $. Expanding the field operator in terms
of the modes and using the commutation relations for the fermionic
annihilation and creation operators, the VEV of the current density is
presented in the form of the mode sum%
\begin{equation}
\langle j^{\mu }(x)\rangle =-\frac{e}{2}\sum_{\sigma }\sum_{\kappa
=-,+}\kappa \bar{\psi}_{\sigma }^{(\kappa )}(x)\gamma ^{\mu }\psi _{\sigma
}^{(\kappa )}(x),  \label{VEVj1}
\end{equation}%
where the terms with $\kappa =+$ and $\kappa =-$ correspond to the
contributions of the positive and negative energy modes.

The structure of the mode functions $\psi _{\sigma }^{(\kappa )}(x)$ is
similar to that discussed in \cite{Bell16T}. They are specified by the
quantum numbers $(\gamma ,j)$, where $j=\pm 1/2,\pm 3/2,\ldots $ is the
total angular momentum and the radial quantum number $\gamma $ determines
the energy of the corresponding mode $\kappa E$, with $E=\sqrt{\gamma
^{2}+m^{2}}$. Introducing the notation
\begin{equation}
\alpha =\chi +eA/q=\chi -e\Phi /(2\pi ),  \label{alfa}
\end{equation}%
the mode functions are presented in the form
\begin{equation}
\psi _{\sigma }^{(\kappa )}(x)=C_{\kappa }e^{iq(j+\chi )\phi -\kappa
iEt}\left(
\begin{array}{c}
g_{\beta _{j},\beta _{j}}(\gamma a,\gamma r)e^{-iq\phi /2} \\
\epsilon _{j}\frac{\gamma e^{iq\phi /2}}{\kappa E+sm}g_{\beta _{j},\beta
_{j}+\epsilon _{j}}(\gamma a,\gamma r)%
\end{array}%
\right) \ ,  \label{psie}
\end{equation}%
where $\epsilon _{j}=1$ for $j>-\alpha $ and $\epsilon _{j}=-1$ for $%
j<-\alpha $,
\begin{equation}
\beta _{j}=q|j+\alpha |-\epsilon _{j}/2.  \label{betj}
\end{equation}%
Note that the part $e\Phi /(2\pi )$ in (\ref{alfa}) is the ratio of the
magnetic flux threading the ring to the flux quantum $\Phi _{0}=2\pi /e$.
The functions $g_{\beta _{j},\nu }(\gamma a,\gamma r)$ of the radial
coordinate $r$, with the orders $\nu =\beta _{j}$ and $\nu =\beta
_{j}+\epsilon _{j}$, is expressed in terms of the Bessel and Neumann
functions as:%
\begin{equation}
g_{\beta _{j},\nu }(\gamma a,\gamma r)=Y_{\beta _{j}}^{(a)}(\gamma a)J_{\nu
}(\gamma r)-J_{\beta _{j}}^{(a)}(\gamma a)Y_{\nu }(\gamma r)\ .  \label{gbet}
\end{equation}%
For the Bessel and Neumann functions we use the notation%
\begin{eqnarray}
f_{\beta _{j}}^{(u)}\left( x\right) &=&xf_{\beta _{j}}^{\prime }\left(
x\right) +[\lambda _{u}n_{u}(\kappa \sqrt{x^{2}+m_{u}^{2}}+sm_{u})-\epsilon
_{j}\beta _{j}]f_{\beta _{j}}\left( x\right)  \notag \\
&=&\lambda _{u}n_{u}(\kappa \sqrt{x^{2}+m_{u}^{2}}+sm_{u})f_{\beta
_{j}}\left( x\right) -\epsilon _{j}xf_{\beta _{j}+\epsilon _{j}}\left(
x\right) ,  \label{fbar}
\end{eqnarray}%
with $u=a,b$, $f=J,Y$, and $m_{u}=mu$. When the parameter $\alpha $ is equal
to an half-integer, the modes with $j\neq -\alpha $ are still given by (\ref%
{psie}). In this case there is a special mode with $j=-\alpha $ which is
separately discussed in appendix \ref{sec:Sp}.

The coefficients of the linear combination of the cylinder functions in (\ref%
{gbet}) are obtained from the boundary condition (\ref{BCs}) at $r=a$. The
further imposition of the boundary condition at $r=b$ determines the
eigenvalues of the quantum number $\gamma $ as roots of the equation%
\begin{equation}
C_{\beta _{j}}(b/a,\gamma a)\equiv J_{\beta _{j}}^{(a)}\left( \gamma
a\right) Y_{\beta _{j}}^{(b)}\left( \gamma b\right) -J_{\beta
_{j}}^{(b)}\left( \gamma b\right) Y_{\beta _{j}}^{(a)}\left( \gamma a\right)
=0.  \label{EigEq}
\end{equation}%
We will denote by $z_{l}$, $l=1,2,\ldots $, the positive solutions of this
equation with respect to $\gamma a$, assuming that $z_{l}<z_{l+1}$. The
eigenvalues of $\gamma $ are expressed as $\gamma =\gamma _{l}=z_{l}/a$.
Hence, the mode functions are specified by the set of discrete quantum
numbers $\sigma =(l,j)$. The energies of the positive and negative energy
modes are given as $E_{\kappa }=\kappa E$ with $E=\sqrt{\gamma _{l}^{2}+m^{2}%
}$. For a given quantum number $j$, the equations (\ref{EigEq}) for the
eigenvalues $\gamma _{l}$ of the positive and negative energy modes differ
by the change of the energy sign. As it will be discussed in appendix \ref%
{sec:App1}, depending on the set of the parameters $(s,\lambda _{a},\lambda
_{b})$, purely imaginary solutions of the equation (\ref{EigEq}) may
present. For all these solutions $\gamma ^{2}+m^{2}\geq 0$ and the vacuum
state is stable. For a massless field the confinement of the field, in
general, induces an energy gap that depends on the geometrical
characteristics of the ring. The controllable energy gap plays an important
role in graphene ribbons. For large values of $\gamma a\gg 1$ we can use in (%
\ref{EigEq}) the asymptotic expressions of cylinder functions for large
arguments. In the case $\lambda _{a}=-\lambda _{b}$, to the leading order,
the equation of the modes is reduced to $\sin \left[ \left( b-a\right)
\gamma \right] =0$ with $\gamma _{l}\approx \pi l/(b-a)$ for large $l$. For $%
\lambda _{a}=\lambda _{b}$ and $\gamma a\gg 1$ from (\ref{EigEq}) we get $%
s(m/\gamma )\sin x+\cos x=0$ with $x=\left( b-a\right) \gamma $. For $\gamma
\gg m$ this gives $\gamma _{l}\approx \pi (l+1/2)/(b-a)$.

Let us present the parameter $\alpha $ from (\ref{alfa}) in the form
\begin{equation}
\alpha =\alpha _{0}+n_{0},\quad |\alpha _{0}|\leq 1/2,  \label{alpha}
\end{equation}%
where $n_{0}$ is an integer. Redefining $j\rightarrow j+n_{0}$, we see that
the solutions $z_{l}$ are functions of $b/a,\alpha _{0},j,s,\lambda
_{u},\kappa $: $z_{l}=z_{l}(b/a,s,\lambda _{u},j,\alpha _{0},\kappa )$. By
taking into account (\ref{fbar}) it can be seen that the function $f_{\beta
_{j}}^{(u)}\left( x\right) $ for the set $(j,\alpha _{0},\kappa )$ coincides
with the function $f_{\beta _{j}}^{(u)}\left( x\right) $ for the set $%
(-j,-\alpha _{0},-\kappa )$ up to the coefficient $\lambda _{u}n_{u}(\kappa
\sqrt{x^{2}+m_{u}^{2}}+sm_{u})/x$. From here it follows that $%
z_{l}(b/a,s,\lambda _{u},j,\alpha _{0},\kappa )=z_{l}(b/a,s,\lambda
_{u},-j,-\alpha _{0},-\kappa )$. In particular, this means that the negative
energy solutions for the set $(j,\alpha _{0})$ coincide with the positive
energy solutions for $(-j,-\alpha _{0})$. Another relation between the roots
for different sets of parameters directly follows from the definition (\ref%
{fbar}): $z_{l}(b/a,s,\lambda _{u},j,\alpha _{0},\kappa
)=z_{l}(b/a,-s,-\lambda _{u},j,\alpha _{0},-\kappa )$. Combining this with
the previous relation we get $z_{l}(b/a,s,\lambda _{u},j,\alpha _{0},\kappa
)=z_{l}(b/a,-s,-\lambda _{u},-j,-\alpha _{0},\kappa )$.

To complete the specification of the fermionic modes it remains to determine
the normalization coefficient $C_{\kappa }$ in (\ref{psie}). It is obtained
from the standard orthonormalization condition
\begin{equation}
\int_{a}^{b}dr\int_{0}^{\phi _{0}}d\phi \,r\psi _{\sigma }^{(\kappa )\dagger
}(x)\psi _{\sigma ^{\prime }}^{(\kappa )}(x)=\delta _{jj^{\prime }}\ \delta
_{ll^{\prime }},  \label{Norm}
\end{equation}%
for fermionic fields. The radial integral involving the square of the
cylinder functions $g_{\beta _{j},\nu }(\gamma a,\gamma r)$ is evaluated by
using the result from \cite{Prud86}. This leads to the following expression%
\begin{equation}
\left\vert C_{\kappa }\right\vert ^{2}=\frac{\pi qz_{l}}{16a^{2}}\frac{%
E+\kappa sm}{E}T_{\beta _{j}}^{ab}(z_{l}),  \label{C+}
\end{equation}%
where we have defined the function%
\begin{equation}
T_{\beta _{j}}^{ab}(z)=\frac{z}{E+\kappa sm}\left[ \frac{B_{b}J_{\beta
_{j}}^{(a)2}\left( z\right) }{J_{\beta _{j}}^{(b)2}\left( zb/a\right) }-B_{a}%
\right] ^{-1},  \label{Tab}
\end{equation}%
with
\begin{equation}
B_{u}=u^{2}\left[ E-\frac{\kappa \lambda _{u}n_{u}}{u}\left( \frac{E-\kappa
sm}{2E}+\epsilon _{j}\beta _{j}\right) \right] .  \label{Bu}
\end{equation}%
In deriving (\ref{C+}) we have used the relations
\begin{eqnarray}
g_{\beta _{j},\beta _{j}}(\gamma a,\gamma u) &=&\frac{2}{\pi }\frac{J_{\beta
_{j}}^{(a)}(\gamma a)}{J_{\beta _{j}}^{(u)}(\gamma u)},  \notag \\
g_{\beta _{j},\beta _{j}+\epsilon _{j}}(\gamma a,\gamma u) &=&\frac{%
2\epsilon _{j}}{\pi \gamma }\lambda _{u}n_{u}(\kappa E+sm)\frac{J_{\beta
_{j}}^{(a)}(\gamma a)}{J_{\beta _{j}}^{(u)}(\gamma u)},  \label{Relg}
\end{eqnarray}%
with $u=a,b$.

The model under consideration is specified by the set of parameters $(\chi
,A)$. The first one determines the phase in the periodicity condition in the
azimuthal direction and the second one determines the magnetic flux enclosed
by the ring. These parameters are not separately gauge invariant. Under the
gauge transformation $A_{\mu }=A_{\mu }^{\prime }+\partial _{\mu }\omega $, $%
\psi (x)=\psi ^{\prime }(x)e^{-ie\omega }$, with the function $\omega
=b_{\mu }x^{\mu }$, a new set is given by $(\chi ^{\prime },A^{\prime
})=(\chi +eb_{2}/q,A-b_{2})$. However, the parameter $\alpha $, defined by (%
\ref{alfa}), is gauge invariant. In particular, in the gauge with $%
b_{2}=-q\chi /e$ the fermionic field is periodic in the azimuthal direction
and the phase $\chi $ is interpreted in terms of a fictitious magnetic flux $%
-2\pi \chi /e=-\chi \Phi _{0}$. In this sense, the parameter $\alpha $ can
be considered as the ratio of the effective magnetic flux to the flux
quantum.

\section{VEVs of the charge and current densities}

\label{sec:jmu}

\subsection{Mode sum}

In this section we evaluate the VEVs of the charge and current densities on
conical rings. First we assume that all the solutions of the eigenvalue
equation (\ref{EigEq}) are real. The modifications in the evaluation
procedure required by the presence of imaginary roots are described in
appendix \ref{sec:App1}. Having specified the complete set of mode functions
(\ref{psie}), for the VEV (\ref{VEVj1}) one finds the representation%
\begin{equation}
\langle j^{\mu }\rangle =-\frac{\pi eq}{32a^{2}}\sum_{j}\sum_{\kappa =\pm
}\,\kappa \sum_{l=1}^{\infty }T_{\beta _{j}}^{ab}(z_{l})w_{\mu ,\beta
_{j}}(z_{l}),  \label{jmu}
\end{equation}%
where in the summation over $j$ one has $j=\pm 1/2,\pm 3/2,\ldots $. Here
for the charge and azimuthal current densities we have defined the functions%
\begin{eqnarray}
w_{0,\beta _{j}}(z) &=&\frac{z}{E}\left[ \left( E+\kappa sm\right) g_{\beta
_{j},\beta _{j}}^{2}(z,zr/a)+\left( E-\kappa sm\right) g_{\beta _{j},\beta
_{j}+\epsilon _{j}}^{2}(z,zr/a)\right] ,  \notag \\
w_{2,\beta _{j}}(z) &=&2\frac{\kappa \epsilon _{j}z^{2}}{arE}g_{\beta
_{j},\beta _{j}}(z,zr/a)g_{\beta _{j},\beta _{j}+\epsilon _{j}}(z,zr/a),
\label{w2}
\end{eqnarray}%
with $E=\sqrt{z^{2}/a^{2}+m^{2}}$ and $w_{1,\beta _{j}}(z)=0$. The VEV\ of
the radial current density vanishes. Note that the physical component of the
azimuthal current density is given by $\langle j_{\phi }\rangle =r\langle
j^{2}\rangle $.

We can see that under the replacements $\beta _{j}\rightleftarrows \beta
_{j}+\epsilon _{j}$,$\;\kappa \rightarrow -\kappa $ the function (\ref{fbar}%
) transforms as
\begin{equation}
f_{\beta _{j}}^{(u)}\left( u\gamma \right) \rightarrow -\epsilon _{j}\left(
\lambda _{u}n_{u}/u\right) (\kappa E+sm)f_{\beta _{j}}^{(u)}\left( u\gamma
\right) .  \label{Transf}
\end{equation}%
From here it follows that the roots $z_{l}$ of (\ref{EigEq}) are not changed
under those replacements. The same is the case for the product $T_{\beta
_{j}}^{ab}(z_{l})w_{\mu ,\beta _{j}}(z_{l})$ in (\ref{jmu}). But the
replacements $\beta _{j}\rightleftarrows \beta _{j}+\epsilon _{j}$ are
equivalent to the change $(j,\alpha )\rightarrow (-j,-\alpha )$. Hence, we
conclude that the VEVs (\ref{jmu}) are odd periodic functions of the
parameter $\alpha $ with the period 1. This implies periodicity with respect
to the enclosed magnetic flux with the period of the flux quantum. Of
course, this is the well known feature for Aharonov-Bohm type effects.

For half-integer values of the parameter $\alpha $ the contribution of the
modes $j\neq -\alpha $ to the VEV $\langle j^{\mu }\rangle $ is still given
by expression (\ref{jmu}) and the contribution coming from the special mode $%
j=-\alpha $ is investigated in appendix \ref{sec:Sp}. Redefining the
summation variable $j$ in (\ref{jmu}), it is sufficient to consider the
values $\alpha =\pm 1/2$. For definiteness consider the case $\alpha =1/2$.
Let us present the series over $j$ in (\ref{jmu2}) as $\sum_{j}\kappa
f(\beta _{j},\beta _{j}+\epsilon _{j},\kappa )$ with $j\neq -1/2$. In the
part over the negative values $j$ we pass to a new summation variable, $%
j\rightarrow -j-1$. This transforms the series to the form $\sum_{j>0}\kappa %
\left[ f(\beta _{j},\beta _{j}+\epsilon _{j},\kappa )+f(\beta _{j}+\epsilon
_{j},\beta _{j},\kappa )\right] $. But as it has been explained above $%
f(\beta _{j}+\epsilon _{j},\beta _{j},\kappa )=f(\beta _{j},\beta
_{j}+\epsilon _{j},-\kappa )$ and the expression under the summation sign is
an odd function of $\kappa $. Hence, the contributions from the positive and
negative energy modes cancel each other and the modes with $j\neq -\alpha $
do not contribute to the charge and current densities for half-integer
values of $\alpha $. As it is shown in appendix \ref{sec:Sp} the same is the
case for the contribution of the mode $j=-\alpha $ if $\lambda _{a}=\lambda
_{b}$. For $\lambda _{a}=-\lambda _{b}$ and $j=-\alpha $ the positive
eigenvalues of $\gamma $ are zeros of the function $\sin [\gamma (b-a)]$
and, again, their contribution vanishes. In the case $\lambda _{a}=-\lambda
_{b}$ the only nonzero contribution comes from the zero energy mode and the
corresponding charge density is given by (\ref{j0s10}). Hence, for
half-integer values of the parameter $\alpha $ the charge and current
densities vanish for the boundary conditions with $\lambda _{a}=\lambda _{b}$
and are determined by $\langle j^{0}\rangle =-\lambda _{a}\langle j_{\phi
}\rangle =\langle j^{0}\rangle _{\mathrm{(s)}}^{(0)}$, with $\langle
j^{0}\rangle _{\mathrm{(s)}}^{(0)}$ from (\ref{j0s10}), for $\lambda
_{a}=-\lambda _{b}$.

Returning to the general case for $\alpha $ and by using the relations (\ref%
{Relg}), for the charge density on the ring edges one finds%
\begin{equation}
\langle j^{0}\rangle _{r=u}=-\frac{eq}{4\pi a}\sum_{j}\sum_{\kappa =\pm
}\,\kappa \sum_{l=1}^{\infty }\gamma _{l}T_{\beta _{j}}^{ab}(z_{l})\frac{%
E+\kappa sm}{E}\frac{J_{\beta _{j}}^{(a)2}(\gamma a)}{J_{\beta
_{j}}^{(u)2}(\gamma u)}.  \label{j0u}
\end{equation}%
The azimuthal current density on the edge is related to the corresponding
charge density by the simple formula%
\begin{equation}
\langle j_{\phi }\rangle _{r=u}=\lambda _{u}n_{u}\langle j^{0}\rangle _{r=u}.
\label{j2u}
\end{equation}%
For planar rings this relation in the case $\lambda _{u}=1$ has been already
mentioned in \cite{Bell16}.

\subsection{Integral representation}

The representation (\ref{jmu}) has two disadvantages: the roots $z_{l}$ are
given implicitly, as zeros of the function (\ref{EigEq}), and the terms with
large $l$ are highly oscillatory. Both of these difficulties can be overcome
by making use of the summation formula \cite{Saha08book} (see also \cite%
{Beze06})
\begin{eqnarray}
\sum_{l=1}^{\infty }w(z_{l})T_{\beta _{j}}^{ab}(z_{l}) &=&\frac{4}{\pi ^{2}}%
\int_{0}^{\infty }{dz\,}\frac{w(z)}{J_{\beta _{j}}^{(a)2}\left( z\right)
+Y_{\beta _{j}}^{(a)2}\left( z\right) }-\frac{2}{\pi }\underset{z=0}{\mathrm{%
Res}}\left[ \frac{w(z)H_{\beta _{j}}^{(1b)}\left( zb/a\right) }{C_{\beta
_{j}}(b/a,z)H_{\beta _{j}}^{(1a)}\left( z\right) }\right]  \notag \\
&&-\frac{1}{\pi }\int_{0}^{\infty }dz\,\sum_{p=+,-}\frac{w(ze^{pi\pi
/2})K_{\beta _{j}}^{(bp)}(\eta z)/K_{\beta _{j}}^{(ap)}(z)}{K_{\beta
_{j}}^{(ap)}\left( z\right) I_{\beta _{j}}^{(bp)}\left( \eta z\right)
-I_{\beta _{j}}^{(ap)}(z)K_{\beta _{j}}^{(bp)}(\eta z)}.  \label{Sum}
\end{eqnarray}%
Here we use the notation (\ref{fbar}) for the Hankel functions $H_{\nu
}^{(1,2)}(x)$ and the notation
\begin{equation}
f_{\beta _{j}}^{(up)}(x)=xf_{\beta _{j}}^{\prime }\left( x\right) +\left\{
\lambda _{u}n_{u}\left[ \kappa \sqrt{\left( xe^{p\pi i/2}\right)
^{2}+m_{u}^{2}}+sm_{u}\right] -\epsilon _{j}\beta _{j}\right\} f_{\beta
_{j}}\left( x\right) ,  \label{fjp}
\end{equation}%
for the modified Bessel functions $I_{\nu }(x)$ and $K_{\nu }(x)$. The
conditions on the function $w(z)$, analytic in the right-half plane $\mathrm{%
Re}\,z>0$, are formulated in \cite{Saha08book}. On the imaginary axis the
function $w(z)$ may have branch points. The square root $\sqrt{\left(
ze^{p\pi i/2}\right) ^{2}+m_{u}^{2}}$ in (\ref{fjp}) is understood as $\sqrt{%
m_{u}^{2}-z^{2}}$, for $z<m_{u}$, and as $pi\sqrt{z^{2}-m_{u}^{2}}$ for $%
z>m_{u}$. From here it follows that $f_{\beta _{j}}^{(u+)}(z)=f_{\beta
_{j}}^{(u-)}(z)$ for $z<m_{u}$.

For the series in (\ref{jmu}) one has $w(z)=w_{\mu ,\beta _{j}}(z)$. The
functions $w_{\mu ,\beta _{j}}(z)$ have branch points $z=\pm im_{a}$ on the
imaginary axis and obey the relation $w_{\mu ,\beta _{j}}(ze^{-\pi
i/2})=-w_{\mu ,\beta _{j}}(ze^{\pi i/2})$ for $z<m_{a}$. By using these
properties we can see that the positive and negative energy modes give the
same contributions to the VEVs of the charge and current densities and they
are presented as
\begin{equation}
\langle j^{\mu }\rangle =\langle j^{\mu }\rangle _{a}+\frac{eq}{2\pi ^{2}}%
\sum_{j}\,\int_{m}^{\infty }dx\,\frac{x}{\sqrt{x^{2}-m^{2}}}\mathrm{Re}\left[
\frac{V_{\mu ,\beta _{j}}^{(a)}(ax,rx)K_{\beta _{j}}^{(b)}(bx)/K_{\beta
_{j}}^{(a)}(ax)}{K_{\beta _{j}}^{(a)}\left( ax\right) I_{\beta
_{j}}^{(b)}\left( bx\right) -I_{\beta _{j}}^{(a)}(ax)K_{\beta _{j}}^{(b)}(bx)%
}\right] ,  \label{jmu2}
\end{equation}%
where%
\begin{eqnarray}
V_{0,\beta _{j}}^{(u)}(ux,rx) &=&\left( sm+i\sqrt{x^{2}-m^{2}}\right)
G_{\beta _{j},\beta _{j}}^{(u)2}(ux,rx)+\left( sm-i\sqrt{x^{2}-m^{2}}\right)
G_{\beta _{j},\beta _{j}+\epsilon _{j}}^{(u)2}(ux,rx),  \notag \\
V_{2,\beta _{j}}^{(u)}(ux,rx) &=&-\frac{2x}{r}G_{\beta _{j},\beta
_{j}}^{(u)}(ux,rx)G_{\beta _{j},\beta _{j}+\epsilon _{j}}^{(u)}(ux,rx),
\label{V2a}
\end{eqnarray}%
with $u=a,b$. The functions in the right-hand sides of (\ref{V2a}) are
defined by
\begin{equation}
G_{\beta _{j},\mu }^{(u)}(x,y)=K_{\beta _{j}}^{(u)}\left( x\right) I_{\mu
}\left( y\right) -(-1)^{\mu -\beta _{j}}I_{\beta _{j}}^{(u)}\left( x\right)
K_{\mu }\left( y\right) ,  \label{Gen}
\end{equation}%
and for the modified Bessel functions $f_{\nu }\left( z\right) =I_{\nu
}\left( z\right) ,K_{\nu }\left( z\right) $ we use the notation%
\begin{eqnarray}
f_{\beta _{j}}^{(u)}(z) &=&zf_{\beta _{j}}^{\prime }\left( z\right) +\left[
\lambda _{u}n_{u}\left( i\sqrt{z^{2}-m_{u}^{2}}+sm_{u}\right) -\epsilon
_{j}\beta _{j}\right] f_{\beta _{j}}\left( z\right)  \notag \\
&=&\delta _{f}zf_{\beta _{j}+\epsilon _{j}}\left( z\right) +\lambda
_{u}n_{u}(i\sqrt{z^{2}-m_{u}^{2}}+sm_{u})f_{\beta _{j}}\left( z\right) ,
\label{fnu}
\end{eqnarray}%
where $\delta _{I}=1$, $\delta _{K}=-1$, and $u=a,b$. The expressions for
the VEVs of the charge and current densities contain a summation over $j$
that enters in the formulas through $\beta _{j}$ defined as (\ref{betj}).
Redefining the summation variable $j\rightarrow j+n_{0}$, with $n_{0}$
defined by (\ref{alpha}), we see that the VEVs do not depend on $n_{0}$ and
only the fractional part of $\alpha $ is physically relevant. Recall that in
deriving (\ref{jmu2}) we have assumed that all the roots of the equation (%
\ref{EigEq}) are real. In appendix \ref{sec:App1} it is shown that the
representation (\ref{jmu2}) is valid also in the presence of imaginary roots
corresponding to the bound states.

In (\ref{jmu2}), the part $\langle j^{\mu }\rangle _{a}$ comes from the
first term in the right-hand side of (\ref{Sum}) and is given by the
expression
\begin{equation}
\langle j^{\mu }\rangle _{a}=-\frac{eq}{8\pi a^{2}}\sum_{j}\sum_{\kappa =\pm
1}\,\int_{0}^{\infty }{dz\,}\frac{\kappa w_{\mu ,\beta _{j}}(z)}{J_{\beta
_{j}}^{(a)2}\left( z\right) +Y_{\beta _{j}}^{(a)2}\left( z\right) }.
\label{jmua}
\end{equation}%
For its physical interpretation we note that the last term in (\ref{jmu2})
tends to zero in the limit $b\rightarrow \infty $. This shows that (\ref%
{jmua}) corresponds to the VEV in the region $r\geq a$ for a cone with a
single edge $r=a$. By using the identity%
\begin{equation}
\frac{g_{\beta _{j},\nu }(z,y)g_{\beta _{j},\rho }(z,y)}{J_{\beta
_{j}}^{(a)2}(x)+Y_{\beta _{j}}^{(a)2}(x)}=J_{\nu }(y)J_{\rho }(y)-\frac{1}{2}%
\sum_{l=1,2}\frac{J_{\beta _{j}}^{(a)}(z)}{H_{\beta _{j}}^{(al)}(z)}H_{\nu
}^{(l)}(y)H_{\rho }^{(l)}(y).  \label{Id1}
\end{equation}%
with $\nu ,\rho =\beta _{j},\beta _{j}+\epsilon _{j}$, it can be further
decomposed as%
\begin{equation}
\langle j^{\mu }\rangle _{a}=\langle j^{\mu }\rangle _{0}+\langle j^{\mu
}\rangle _{a}^{\mathrm{(b)}},  \label{j0adec}
\end{equation}%
where the separate parts come from the first and second terms in the
right-hand side of (\ref{Id1}). For the first part one has%
\begin{equation}
\langle j^{\mu }\rangle _{0}=-\frac{eq}{4\pi }\sum_{j}\,\int_{0}^{\infty
}dx\,\frac{xw_{\mu ,\beta _{j}}^{(0)}(rx)}{\sqrt{x^{2}+m^{2}}},  \label{jmu0}
\end{equation}%
with the functions%
\begin{eqnarray}
w_{0,\beta _{j}}^{(0)}(z) &=&sm\left[ J_{\beta _{j}}^{2}(z)-J_{\beta
_{j}+\epsilon _{j}}^{2}(z)\right] ,  \notag \\
w_{2,\beta _{j}}^{(0)}(z) &=&\frac{2\epsilon _{j}z}{r^{2}}J_{\beta
_{j}}(z)J_{\beta _{j}+\epsilon _{j}}(z),  \label{w02}
\end{eqnarray}%
and $w_{1,\beta _{j}}^{(0)}(z)=0$. In the part $\langle j^{\mu }\rangle
_{a}^{\mathrm{(b)}}$ we rotate the contour of the integration over $z$ by
the angles $\pi /2$ and $-\pi /2$ for the terms with $l=1$ and $l=2$,
respectively. Introducing the modified Bessel functions we get%
\begin{equation}
\langle j^{\mu }\rangle _{a}^{\mathrm{(b)}}=\frac{eq}{2\pi ^{2}}%
\sum_{j}\int_{m}^{\infty }dx\,\frac{x}{\sqrt{x^{2}-m^{2}}}\mathrm{Re}\left[
\frac{I_{\beta _{j}}^{(a)}(ax)}{K_{\beta _{j}}^{(a)}(ax)}W_{\mu ,\beta
_{j}}(rx)\right] ,  \label{jmua2}
\end{equation}%
with the notations (\ref{fnu}) and%
\begin{eqnarray}
W_{0,\beta _{j}}(rx) &=&\left( sm+i\sqrt{x^{2}-m^{2}}\right) K_{\beta
_{j}}^{2}(rx)+\left( sm-i\sqrt{x^{2}-m^{2}}\right) K_{\beta _{j}+\epsilon
_{j}}^{2}(rx),  \notag \\
W_{2,\beta _{j}}(rx) &=&\frac{2x}{r}K_{\beta _{j}}(rx)K_{\beta _{j}+\epsilon
_{j}}(rx).  \label{W2}
\end{eqnarray}%
For the representation $s=1$ and for the boundary condition with $\lambda
_{a}=1$, this expression for a single boundary-induced part coincides with
the one given in \cite{Beze10} (comparing the formulas here with the results
of \cite{Beze10}, the replacements $\alpha \rightarrow -\alpha $ and $\alpha
_{0}\rightarrow -\alpha _{0}$ should be made; this difference is related to
that in \cite{Beze10}, for the evaluation of the VEVs for the geometry with
a single boundary, the analog of the negative-energy mode functions (\ref%
{psie}) was used with $\alpha $ replaced by $-\alpha $). The part $\langle
j^{\mu }\rangle _{0}$ in (\ref{j0adec}) with $0<r<\infty $ corresponds to
the VEV in a conical space without boundaries and the contribution $\langle
j^{\mu }\rangle _{a}^{\mathrm{(b)}}$ is induced in the region $r\geq a$ by
the presence of the edge $r=a$.

Another representation of the VEV (\ref{jmu0}) in the boundary-free conical
geometry for the case $s=1$ is provided in \cite{Beze10}. The parameter $s$
enters in (\ref{jmu0}) as a coefficient in the charge density and the
corresponding generalization is straightforward with the expression
\begin{eqnarray}
&&\langle j^{\mu }\rangle _{0}=-\frac{e}{2\pi r}\Big\{\sideset{}{'}{\sum}%
_{l=1}^{[q/2]}(-1)^{l}\sin (2\pi l\alpha _{0})f_{\mu }\left( 2mr\sin (\pi
l/q)\right)  \notag \\
&&\qquad -\frac{q}{\pi }\int_{0}^{\infty }dy\frac{f_{\mu }\left( 2mr\cosh
y\right) }{\cosh (2qy)-\cos (q\pi )}\sum_{p=\pm 1}p\cos \left[ q\pi \left(
1/2-p\alpha _{0}\right) \right] \cosh \left[ q\left( 1+2p\alpha _{0}\right) y%
\right] \Big\},  \label{jm02}
\end{eqnarray}%
where $[q/2]$ means the integer part of $q/2$, the prime on the summation
sign means that for even $q$ the term with $l=q/2$ should be taken with an
additional coefficient 1/2, and we have introduced the functions
\begin{eqnarray}
f_{0}(z) &=&sme^{-z},  \notag \\
f_{2}(z) &=&2m^{2}(1+z)e^{-z}/z^{2}.  \label{f2z}
\end{eqnarray}%
The boundary-free contributions to the charge density for the fields with $%
s=+1$ and $s=-1$ differ only in sign, whereas the azimuthal current
densities coincide.

We can also further transform the edge-induced contributions to the VEVs.
The dependence on $j$ enters through $\beta _{j}$ and $\beta _{j}+\epsilon
_{j}$ (see (\ref{fnu})). It can be seen that for both the series in (\ref%
{jmu2}) and (\ref{jmua2}) one has%
\begin{equation}
\sum_{j}g(\beta _{j},\beta _{j}+\epsilon _{j})=\sum_{n=0}^{\infty
}\sum_{p=\pm 1}pg(n_{p},n_{p}+1),  \label{sumn}
\end{equation}%
with the notation%
\begin{equation}
n_{p}=q(n+1/2+p\alpha _{0})-1/2.  \label{np}
\end{equation}%
As a consequence, the VEVs are presented in the form%
\begin{eqnarray}
\langle j^{\mu }\rangle &=&\langle j^{\mu }\rangle _{0}+\frac{eq}{2\pi ^{2}}%
\sum_{n=0}^{\infty }\sum_{p=\pm 1}p\,\int_{m}^{\infty }dx\,\frac{x}{\sqrt{%
x^{2}-m^{2}}}\mathrm{Re}\left[ \frac{I_{n_{p}}^{(a)}(ax)}{K_{n_{p}}^{(a)}(ax)%
}W_{\mu ,n_{p}}(rx)\right.  \notag \\
&&\left. +\frac{V_{\mu
,n_{p}}^{(a)}(ax,rx)K_{n_{p}}^{(b)}(bx)/K_{n_{p}}^{(a)}(ax)}{%
K_{n_{p}}^{(a)}\left( ax\right) I_{n_{p}}^{(b)}\left( bx\right)
-I_{n_{p}}^{(a)}(ax)K_{n_{p}}^{(b)}(bx)}\right] ,  \label{jmu3}
\end{eqnarray}%
where the functions $W_{\mu ,n_{p}}(rx)$ and $V_{\mu ,n_{p}}^{(a)}(ax,rx)$
are given by (\ref{W2}) and (\ref{V2a}) with the replacements $\beta
_{j}\rightarrow n_{p}$ and $\epsilon _{j}\rightarrow 1$. The same
replacements should be done in the notation (\ref{fnu}) for the modified
Bessel functions. Namely, in (\ref{jmu3})%
\begin{equation}
f_{n_{p}}^{(u)}(z)=\delta _{f}zf_{n_{p}+1}\left( z\right) +\lambda
_{u}n_{u}(i\sqrt{z^{2}-m_{u}^{2}}+sm_{u})f_{n_{p}}\left( z\right) ,
\label{fnpu}
\end{equation}%
for the functions $f_{\nu }\left( z\right) =I_{\nu }\left( z\right) ,K_{\nu
}\left( z\right) $. Note that the ratio of the combinations of the modified
Bessel functions in (\ref{jmu3}) can be presented in the form%
\begin{equation}
\frac{I_{n_{p}}^{(u)}(z)}{K_{n_{p}}^{(u)}(z)}=\frac{W_{n_{p}}^{(u)}(z)-i%
\lambda _{u}n_{u}\sqrt{1-m_{u}^{2}/z^{2}}}{z\left[ K_{n_{p}+1}^{2}\left(
z\right) +K_{n_{p}}^{2}\left( z\right) \right] -2\lambda
_{u}n_{u}sm_{u}K_{n_{p}}\left( z\right) K_{n_{p}+1}\left( z\right) },
\label{IKrat}
\end{equation}%
where%
\begin{eqnarray}
W_{n_{p}}^{(u)}(z) &=&z\left[ I_{n_{p}}\left( z\right) K_{n_{p}}\left(
z\right) -I_{n_{p}+1}\left( z\right) K_{n_{p}+1}\left( z\right) \right]
\notag \\
&&+\lambda _{u}n_{u}sm_{u}\left[ I_{n_{p}+1}\left( z\right) K_{n_{p}}\left(
z\right) -I_{n_{p}}\left( z\right) K_{n_{p}+1}\left( z\right) \right] .
\label{Wu}
\end{eqnarray}%
Under the replacement of the parameters $\lambda _{u}\rightarrow -\lambda
_{u}$, $s\rightarrow -s$ one has $f_{n_{p}}^{(u)}(z)\rightarrow
f_{n_{p}}^{(u)\ast }(z)$, $W_{\mu ,n_{p}}(rx)\rightarrow (-1)^{1-\mu
/2}W_{\mu ,n_{p}}^{\ast }(rx)$, and $V_{\mu ,n_{p}}^{(a)}(ax,rx)\rightarrow
(-1)^{1-\mu /2}V_{\mu ,n_{p}}^{(a)\ast }(ax,rx)$. From here it follows that
for the fields with the parameters $(\lambda _{u},s)$ and $(-\lambda
_{u},-s) $ the VEVs of the charge densities differ in sign, whereas the
current densities are the same. The expression (\ref{jmu3}) explicitly shows
that both the charge and current densities are odd periodic functions of the
magnetic flux threading the ring with the period equal to the flux quantum.
The periodicity of the physical characteristics in the magnetic flux is a
common feature for the Aharonov-Bohm type effects.

As it has been already mentioned, the part in (\ref{jmu3}) with the second
term in the square brackets tends to zero in the limit $b\rightarrow \infty $%
. For a massive field and for fixed $r$ and $a$, that part decays
exponentially, like $e^{-2bm}$ for $b\rightarrow \infty $. In the case of a
massless field the decay, as a function of $b$, is power law: as $\left(
a/b\right) ^{q(1-2|\alpha _{0}|)+1}$ for the charge density and as $\left(
a/b\right) ^{q(1-2|\alpha _{0}|)+2}$ for the azimuthal current. Once again,
this shows that the contribution (\ref{j0adec}) corresponds to the VEVs
outside a single boundary at $r=a$ and the part with the second term in the
square brackets of (\ref{jmu3}) is induced by the outer boundary.

\subsection{Another representation}

The representation (\ref{jmu2}) for the charge and current densities in the
ring is not symmetric with respect to the inner and outer edges. An
alternative representation, with the extracted outer boundary part is
obtained from (\ref{jmu2}) by making use of the relation%
\begin{eqnarray}
&&(-1)^{\nu -\rho }\frac{I_{\beta _{j}}^{(a)}(ax)}{K_{\beta _{j}}^{(a)}(ax)}%
K_{\nu }(y)K_{\rho }(y)+\frac{K_{\beta _{j}}^{(b)}(bx)}{K_{\beta
_{j}}^{(a)}(ax)}\frac{G_{\beta _{j},\nu }^{(a)}(ax,y)G_{\beta _{j},\rho
}^{(a)}(ax,y)}{K_{\beta _{j}}^{(a)}\left( ax\right) I_{\beta
_{j}}^{(b)}\left( bx\right) -I_{\beta _{j}}^{(a)}(ax)K_{\beta _{j}}^{(b)}(bx)%
}  \notag \\
&&\qquad =\frac{K_{\beta _{j}}^{(b)}(bx)}{I_{\beta _{j}}^{(b)}\left(
bx\right) }I_{\nu }\left( y\right) I_{\rho }\left( y\right) +\frac{I_{\beta
_{j}}^{(a)}(ax)}{I_{\beta _{j}}^{(b)}(bx)}\frac{G_{\beta _{j},\nu
}^{(b)}(bx,y)G_{\beta _{j},\rho }^{(b)}(bx,y)}{K_{\beta _{j}}^{(a)}\left(
ax\right) I_{\beta _{j}}^{(b)}\left( bx\right) -I_{\beta
_{j}}^{(a)}(ax)K_{\beta _{j}}^{(b)}(bx)},  \label{Id3}
\end{eqnarray}%
with $\nu ,\rho =\beta _{j},\beta _{j}+\epsilon _{j}$. The expressions for
the VEVs of the charge and current densities take the form%
\begin{equation}
\langle j^{\mu }\rangle =\langle j^{\mu }\rangle _{b}+\frac{eq}{2\pi ^{2}}%
\sum_{j}\,\int_{m}^{\infty }dx\,\frac{x}{\sqrt{x^{2}-m^{2}}}\mathrm{Re}\left[
\frac{V_{\mu ,\beta _{j}}^{(b)}(bx,rx)I_{\beta _{j}}^{(a)}(ax)/I_{\beta
_{j}}^{(b)}(bx)}{K_{\beta _{j}}^{(a)}\left( ax\right) I_{\beta
_{j}}^{(b)}\left( bx\right) -I_{\beta _{j}}^{(a)}(ax)K_{\beta _{j}}^{(b)}(bx)%
}\right] .  \label{jmu4}
\end{equation}%
where the functions $V_{\mu ,\beta _{j}}^{(b)}(bx,rx)$ are defined by (\ref%
{V2a}) with $u=b$. Here, the first term in the right-hand side is decomposed
as%
\begin{equation}
\langle j^{\mu }\rangle _{b}=\langle j^{\mu }\rangle _{0}+\langle j^{\mu
}\rangle _{b}^{\mathrm{(b)}},  \label{j0b}
\end{equation}%
with%
\begin{equation}
\langle j^{\mu }\rangle _{b}^{\mathrm{(b)}}=\frac{eq}{2\pi ^{2}}%
\sum_{j}\int_{m}^{\infty }dx\,\frac{x}{\sqrt{x^{2}-m^{2}}}\mathrm{Re}\left[
\frac{K_{\beta _{j}}^{(b)}(bx)}{I_{\beta _{j}}^{(b)}(bx)}U_{\mu ,\beta
_{j}}(rx)\right] .  \label{jmub}
\end{equation}%
and with the notations defined in accordance with (\ref{fnpu}). The
functions in the integrand are given by the expressions%
\begin{eqnarray}
U_{0,\beta _{j}}(rx) &=&\left( sm+i\sqrt{x^{2}-m^{2}}\right) I_{\beta
_{j}}^{2}(rx)+\left( sm-i\sqrt{x^{2}-m^{2}}\right) I_{\beta _{j}+\epsilon
_{j}}^{2}(rx),  \notag \\
U_{2,\beta _{j}}(rx) &=&-\frac{2x}{r}I_{\beta _{j}}(rx)I_{\beta
_{j}+\epsilon _{j}}(rx).  \label{U2}
\end{eqnarray}%
For the special case with $s=1$ and $\lambda _{b}=1$ the expression (\ref%
{jmub}) coincides with the corresponding result in \cite{Beze10} (with the
replacement $\alpha \rightarrow -\alpha $) for the VEVs inside a single
circular boundary at $r=b$.

Passing from the summation over $j$ to the summation over $n$ in accordance
with (\ref{sumn}), we obtain the final representation%
\begin{eqnarray}
\langle j^{\mu }\rangle &=&\langle j^{\mu }\rangle _{0}+\frac{eq}{2\pi ^{2}}%
\sum_{n=0}^{\infty }\sum_{p=\pm 1}p\,\int_{m}^{\infty }dx\,\frac{x}{\sqrt{%
x^{2}-m^{2}}}\mathrm{Re}\left[ \frac{K_{n_{p}}^{(b)}(bx)}{I_{n_{p}}^{(b)}(bx)%
}U_{\mu ,n_{p}}(rx)\right.  \notag \\
&&\left. +\frac{V_{\mu
,n_{p}}^{(b)}(bx,rx)I_{n_{p}}^{(a)}(ax)/I_{n_{p}}^{(b)}(bx)}{%
K_{n_{p}}^{(a)}\left( ax\right) I_{n_{p}}^{(b)}\left( bx\right)
-I_{n_{p}}^{(a)}(ax)K_{n_{p}}^{(b)}(bx)}\right] .  \label{jmu5}
\end{eqnarray}%
For the ratio under the sign of the real part in (\ref{jmu5}) we have the
following explicit expression%
\begin{equation}
\frac{K_{n_{p}}^{(u)}(z)}{I_{n_{p}}^{(u)}(z)}=\frac{W_{n_{p}}^{(u)}(z)+i%
\lambda _{u}n_{u}\sqrt{1-m_{u}^{2}/z^{2}}}{z[I_{n_{p}+1}^{2}\left( z\right)
+I_{n_{p}}^{2}\left( z\right) ]+2\lambda _{u}n_{u}sm_{u}I_{n_{p}}\left(
z\right) I_{n_{p}+1}\left( z\right) }.  \label{KI}
\end{equation}%
The denominator in this expression is positive for $z\geqslant m_{u}$.
Relatively simple expressions are obtained for a massless field. In the
limit $a\rightarrow 0$ the second term in the square brackets of (\ref{jmu5}%
) behaves like $a^{q(1-2|\alpha _{0}|)}$ and it vanishes for $|\alpha
_{0}|<1/2$. From here it follows that the part $\langle j^{\mu }\rangle _{b}$
corresponds to the VEV in the region $r\leq b$ for the geometry of a single
boundary at $r=b$ for special case of the boundary condition on the cone
apex. The latter correspond to the imposition of the boundary condition (\ref%
{BCs}) on the circle $r=a$ with the subsequent limiting transition $%
a\rightarrow 0$. The part with the last term in the square brackets of (\ref%
{jmu5}) can be interpreted as the contribution of the inner boundary.

\section{Limiting cases and numerical analysis}

\label{sec:Asymp}

In this section we consider some limiting cases of the general results given
above and present the numerical analysis of the behavior of the charge and
current densities as functions of the parameters of the model. The limiting
transitions $a\rightarrow 0$ and $b\rightarrow \infty $ have been already
discussed in the previous section. We have seen that the contributions to $%
\langle j^{\mu }\rangle $, $\mu =0,2$, induced by adding the second boundary
to the geometry of a single boundary decay as $a^{q(1-2|\alpha _{0}|)}$ for
the limit $a\rightarrow 0$ and as $e^{-2bm}$ for $b\rightarrow \infty $. For
a massless field the contribution of the outer boundary in the limit $%
b\rightarrow \infty $ behaves as $1/b^{q(1-2|\alpha _{0}|)+1+\mu /2}$.

Limiting transition to the geometry of a conical space with a single
boundary at $r=b$, corresponding to $a\rightarrow 0$, can also be seen on
the level of the mode functions and of the eigenvalues for the radial
quantum number $\gamma $. In that limit one has $J_{\beta _{j}}^{(a)}\left(
\gamma a\right) \propto a^{q|j+\alpha |+1/2}$ and $Y_{\beta
_{j}}^{(a)}\left( \gamma a\right) \propto a^{1/2-q|j+\alpha |}$. With these
asymptotics, from (\ref{EigEq}) it follows that if $\alpha $ is not equal to
a half-integer then the eigenvalues of $\gamma $ are roots of the equation $%
J_{\beta _{j}}^{(b)}\left( \gamma b\right) =0$. For the mode functions from (%
\ref{psie}) we get%
\begin{equation}
\psi _{\sigma }^{(\kappa )}(x)=C_{\kappa }^{(0)}e^{iq(j+\chi )\phi -\kappa
iEt}\left(
\begin{array}{c}
J_{\beta _{j}}(\gamma r)e^{-iq\phi /2} \\
\epsilon _{j}\frac{\gamma e^{iq\phi /2}}{\kappa E+sm}J_{\beta _{j}+\epsilon
_{j}}(\gamma r)%
\end{array}%
\right) \ ,  \label{psielim}
\end{equation}%
with the normalization coefficient
\begin{equation}
\left\vert C_{\kappa }^{(0)}\right\vert ^{2}=\frac{q\gamma ^{2}}{2\pi }\frac{%
J_{\beta _{j}}^{-2}\left( \gamma b\right) }{2bE\left( bE-\kappa \lambda
_{b}\epsilon _{j}\beta _{j}\right) -\kappa \lambda _{b}b(E-\kappa sm)}.
\label{Ckapb}
\end{equation}%
For $\chi =0$ and $\lambda _{b}=1$ this result coincides with that given in
\cite{Bell16T}.

As it has been shown above, for $\lambda _{a}=\lambda _{b}$ the charge and
current densities vanish for half-integer values of $\alpha $ corresponding
to $\alpha _{0}=\pm 1/2$. That property can also be seen on the base of the
representation (\ref{jmu3}). Let us consider the case $\alpha
_{0}\rightarrow 1/2$. In (\ref{jmu3}), for the part with $p=+1$ we pass to
the summation over $n^{\prime }=n+1$ and then redefine $n^{\prime
}\rightarrow n$. All the terms with $n=1,2,\ldots $ in the parts with $p=+1$
and $p=-1$ cancel each other and the only nonzero contribution comes from
the $n=0$ term in the part with $p=-1$. The expressions for the VEVs of the
charge and current densities are obtained from (\ref{jmu3}) omitting the
summation over $n$ and taking $n_{p}=-1/2$. By using the expressions for the
functions $I_{\pm 1/2}(x)$ and $K_{1/2}(x)$, it can be seen that%
\begin{equation}
\frac{axK_{-1/2}^{(b)}(bx)/K_{-1/2}^{(a)}(ax)}{K_{-1/2}^{(a)}\left(
ax\right) I_{-1/2}^{(b)}\left( bx\right)
-I_{-1/2}^{(a)}(ax)K_{-1/2}^{(b)}(bx)}=-\frac{1+ism/\sqrt{x^{2}-m^{2}}}{%
\frac{sm+\lambda _{a}x}{sm-\lambda _{b}x}e^{2\left( b-a\right) x}-1},
\label{relalf12}
\end{equation}%
and
\begin{eqnarray}
V_{\mu ,-1/2}^{(a)}(ax,rx) &=&\left( \frac{x}{2r}\right) ^{\mu /2}\frac{2a}{%
rx^{2}}(sm+i\sqrt{x^{2}-m^{2}})\left\{ \left( 1-\frac{\mu }{2}\right) \left(
x^{2}-m^{2}\right) \right.  \notag \\
&&\left. +\left( \frac{sm}{2}\right) ^{1-\frac{\mu }{2}}\left[ \left(
sm-\lambda _{a}x\right) e^{2x(a-r)}+(-1)^{\frac{\mu }{2}}\left( sm+\lambda
_{a}x\right) e^{-2x(a-r)}\right] \right\} ,  \label{rel2alf12}
\end{eqnarray}%
for $\mu =0,2$. From these relations it follows that the real part of the
last term in (\ref{jmu3}) is zero and, hence, $\lim_{\alpha _{0}\rightarrow
1/2}\langle j^{\mu }\rangle =\lim_{\alpha _{0}\rightarrow 1/2}\langle j^{\mu
}\rangle _{a}$, where $\langle j^{\mu }\rangle _{a}$ is decomposed as (\ref%
{j0adec}). In the part $\langle j^{\mu }\rangle _{a}^{\mathrm{(b)}}$ we use
the relation
\begin{equation}
\frac{I_{-1/2}^{(a)}(ax)}{K_{-1/2}^{(a)}(ax)}=\frac{1}{\pi }\left( \frac{i%
\sqrt{x^{2}-m^{2}}}{\lambda _{a}x+sm}e^{2ax}+1\right) ,  \label{rel3alf12}
\end{equation}%
to see that%
\begin{equation}
\lim_{\alpha _{0}\rightarrow 1/2}\langle j^{\mu }\rangle =\lim_{\alpha
_{0}\rightarrow 1/2}\langle j^{\mu }\rangle _{0}-\frac{esmq}{2\pi ^{2}r}%
\left( \frac{s}{r}\right) ^{\mu /2}K_{\mu /2}(2mr),  \label{limalf12}
\end{equation}%
with $\mu =0,2$. Now, by using (\ref{jm02}), we can see that the limiting
value $\lim_{\alpha _{0}\rightarrow 1/2}\langle j^{\mu }\rangle _{0}$
exactly cancels the last term in (\ref{limalf12}) and we get $\lim_{\alpha
_{0}\rightarrow 1/2}\langle j^{\mu }\rangle =0$. In particular, for $\lambda
_{a}=\lambda _{b}$ the charge and current densities are continuous function
of the magnetic flux. This is not the case for the VEVs in the boundary-free
conical geometry and also inside a single circular boundary (see also the
discussion in \cite{Beze10}). The VEVs $\langle j^{\mu }\rangle _{0}$ and $%
\langle j^{\mu }\rangle _{b}$ tend to nonzero value in the limit $\alpha
_{0}\rightarrow 1/2$ and the charge and current densities are discontinuous
functions of $\alpha $ at half-integer values of this parameter. Note that
in the case $\lambda _{a}=-\lambda _{b}$ and in the limit $\alpha
_{0}\rightarrow 1/2$ the expression under the Re sign in (\ref{jmu3}) has
pole and its contribution should be appropriately taken into account.
Nonzero limiting values of the charge and current densities for boundary
conditions with $\lambda _{a}=-\lambda _{b}$ are related to that
contribution.

\begin{figure}[tbph]
\begin{center}
\begin{tabular}{cc}
\epsfig{figure=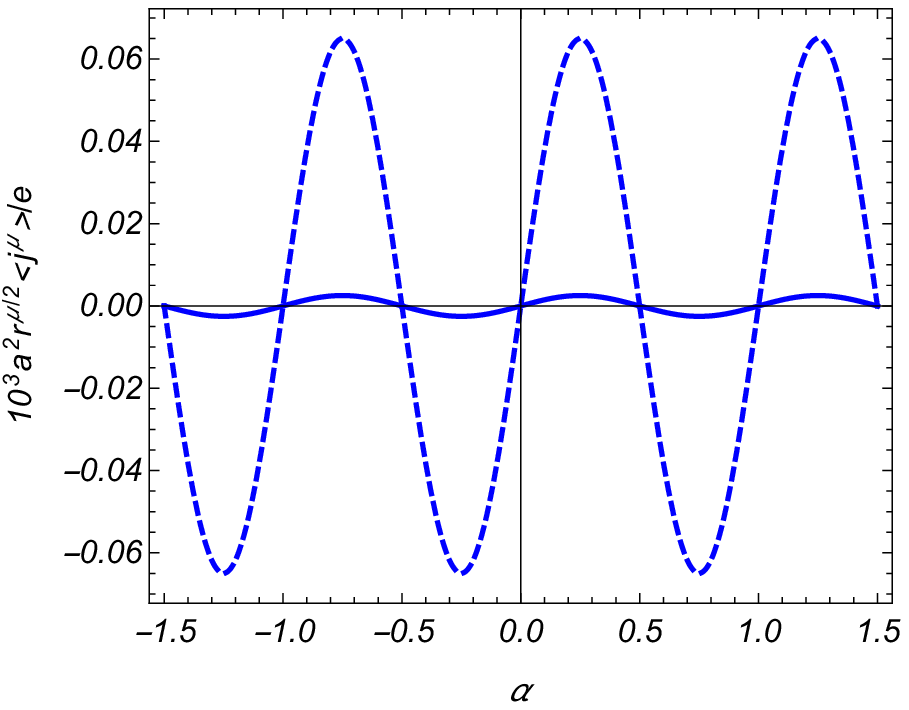,width=7.cm,height=5.5cm} & \quad %
\epsfig{figure=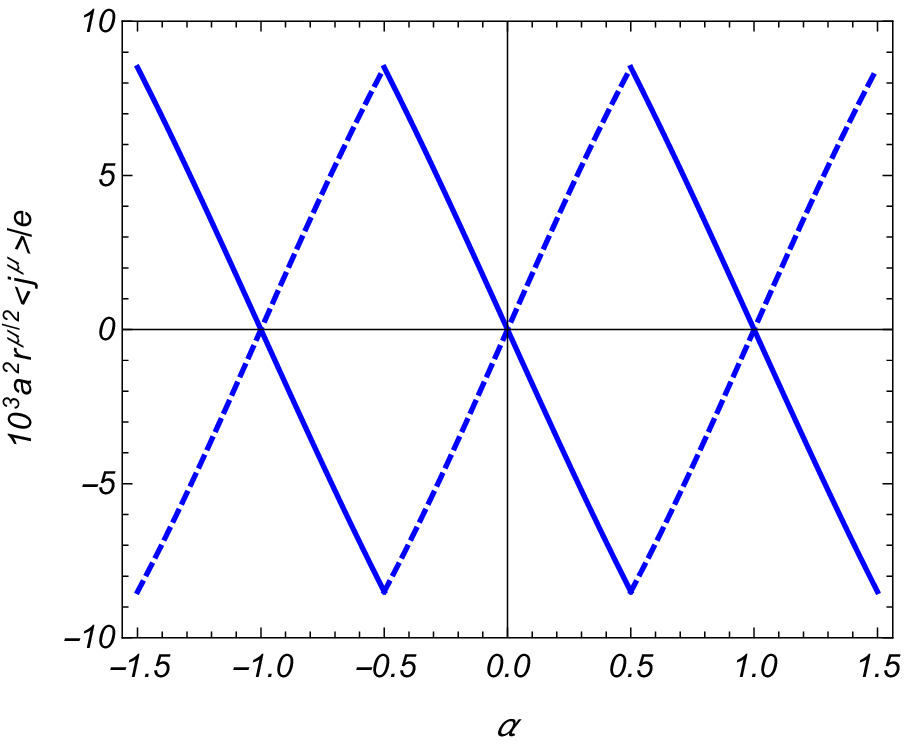,width=7.cm,height=5.5cm}%
\end{tabular}%
\end{center}
\caption{The dependence of the charge (full curves) and current (dashed
curves) densities on the parameter $\protect\alpha $ for the field with $s=1$%
. The graphs are plotted for $q=1.5$, $ma=0.5$, $b/a=4$, $r/a=2$. For the
left panel $\protect\lambda _{a}=\protect\lambda _{b}=1$ and for the right
one $\protect\lambda _{a}=-\protect\lambda _{b}=1$. }
\label{fig2}
\end{figure}

\begin{figure}[tbph]
\begin{center}
\begin{tabular}{cc}
\epsfig{figure=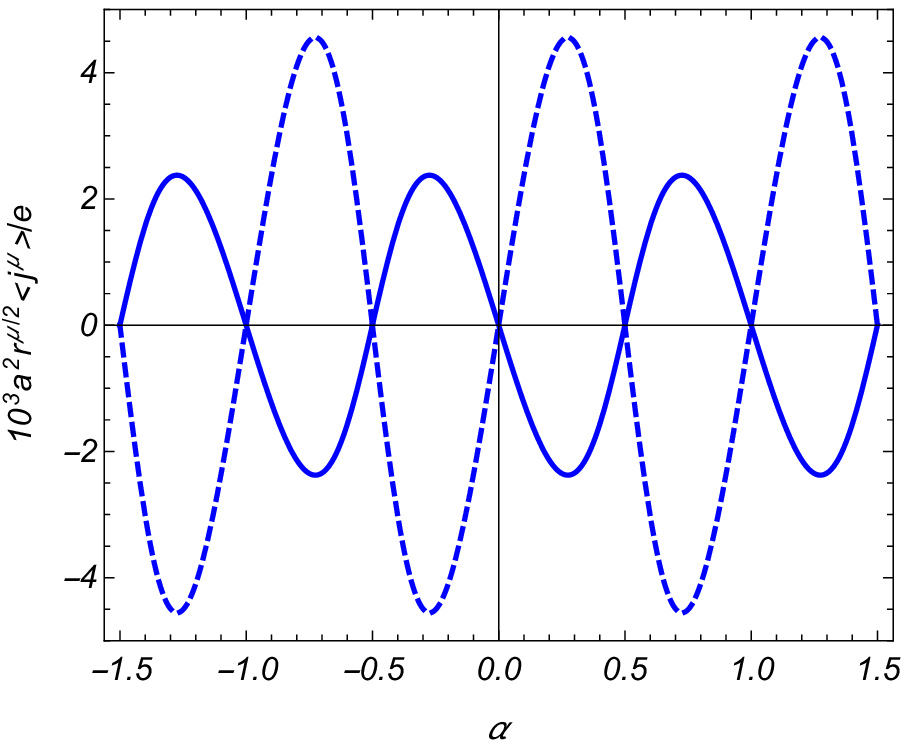,width=7.cm,height=5.5cm} & \quad %
\epsfig{figure=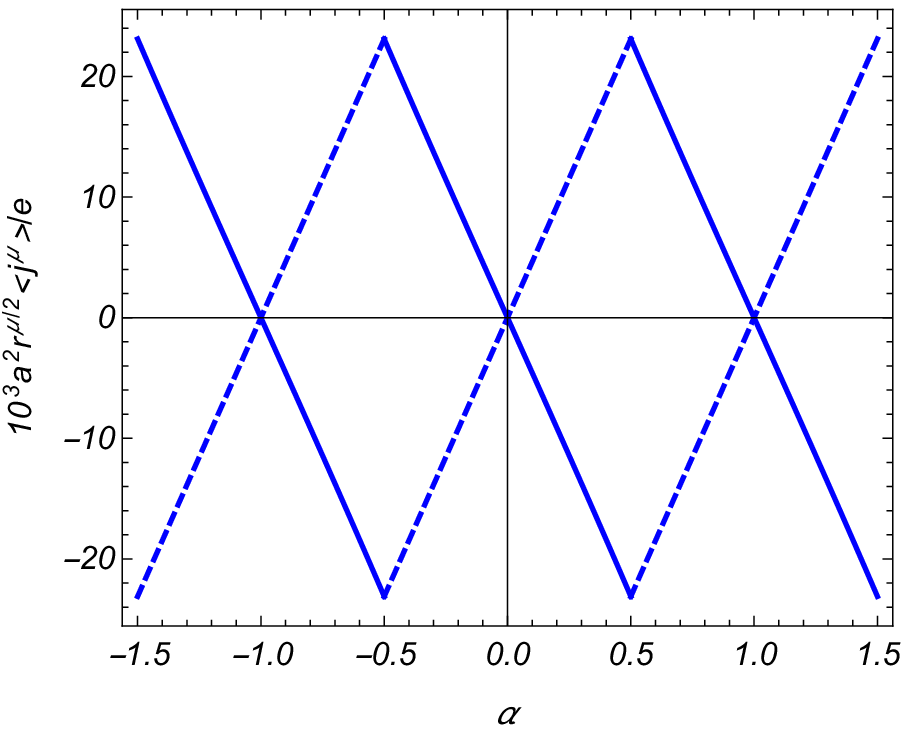,width=7.cm,height=5.5cm}%
\end{tabular}%
\end{center}
\caption{The same as in figure \protect\ref{fig2} for the fermionic field
with $s=-1$.}
\label{fig3}
\end{figure}

The different behavior of the VEVs in the limits $\alpha _{0}\rightarrow \pm
1/2$ for the cases $\lambda _{a}=\lambda _{b}$ and $\lambda _{a}=-\lambda
_{b}$ is seen in figures \ref{fig2} and \ref{fig3}. On those figures we have
plotted the charge (full curves) and current (dashed curves) densities at
the radial point $r/a=2$ as functions of the parameter $\alpha $ for a
conical ring with $q=1.5$, $b/a=4$, and for the mass corresponding to $ma=0.5
$. The figure \ref{fig2} corresponds to the field with $s=1$ and for figure %
\ref{fig3} $s=-1$. The left and right panels on both the figures are plotted
for $\lambda _{a}=\lambda _{b}=1$ and $\lambda _{a}=-\lambda _{b}=1$,
respectively. As is seen from the graphs, the behavior of the VEVs near
half-integer values of $\alpha $ is essentially different for the cases $%
\lambda _{a}=\lambda _{b}$ and $\lambda _{a}=-\lambda _{b}$ (left and right
panels, respectively). In the first case the VEVs vanish at those values
(corresponding to $\alpha _{0}=\pm 1/2$) and they are continuous periodic
functions of the magnetic flux. For $\lambda _{a}=-\lambda _{b}$ the charge
and current densities tend to nonzero limiting values in the limit $\alpha
_{0}\rightarrow \pm 1/2$ and as a consequence of that they are discontinuous
at the half-integer values for $\alpha $. This kind of discontinuities are
present also for persistent currents in mesoscopic normal metal rings. They
appear due to the degeneracy of the energy levels at the corresponding
values of the magnetic flux (see, for example, \cite{Karm07}). As it has
been discussed in appendix \ref{sec:Sp}, in the case $\lambda _{a}=-\lambda
_{b}$ there is a zero energy mode for the angular quantum number $j=-\alpha $
and the nonzero values of the charge and current densities for $\alpha
_{0}=\pm 1/2$ are related to the contribution of that mode. We note that for
the case $\lambda _{a}=-\lambda _{b}$ (right panels) the approximate
relation $\langle j_{\phi }\rangle \approx -\lambda _{a}\langle j^{0}\rangle
$ between the charge and current densities is obeyed to good enough accuracy
for other values of the radial coordinate. This relation is exact for the
zero energy mode. We have also numerically checked the limiting values of
the charge and current densities for the boundary conditions with $\lambda
_{a}=-\lambda _{b}$ obtained from (\ref{jmu3}) when $\alpha _{0}\rightarrow
\pm 1/2$ coincide with the contribution of the zero mode (\ref{j0s10}) for $%
\alpha _{0}=\pm 1/2$.

\begin{figure}[tbph]
\begin{center}
\begin{tabular}{cc}
\epsfig{figure=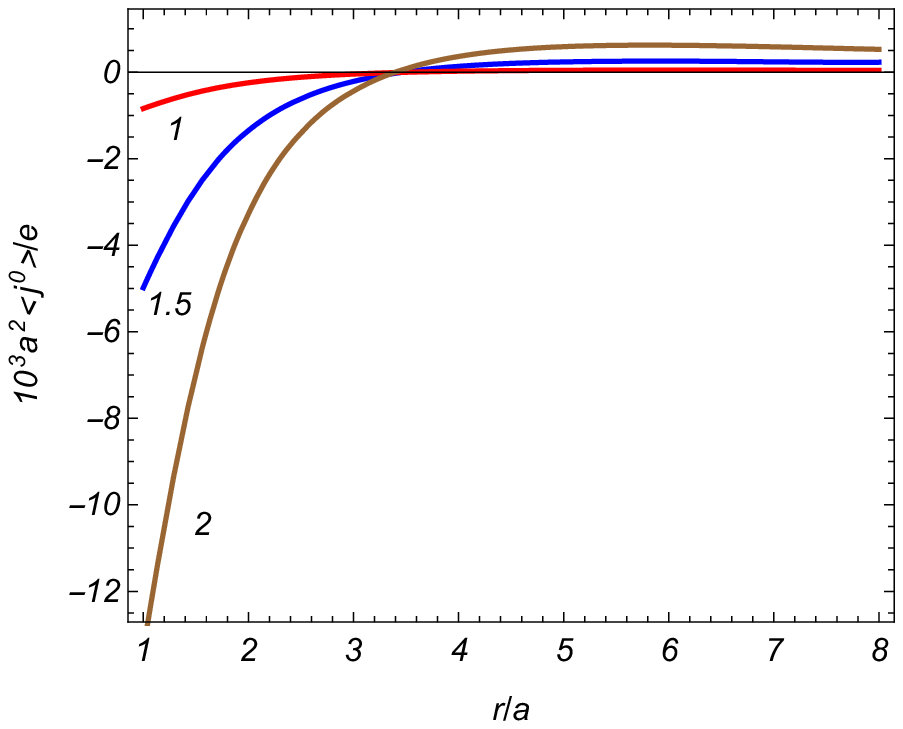,width=7.cm,height=5.5cm} & \quad %
\epsfig{figure=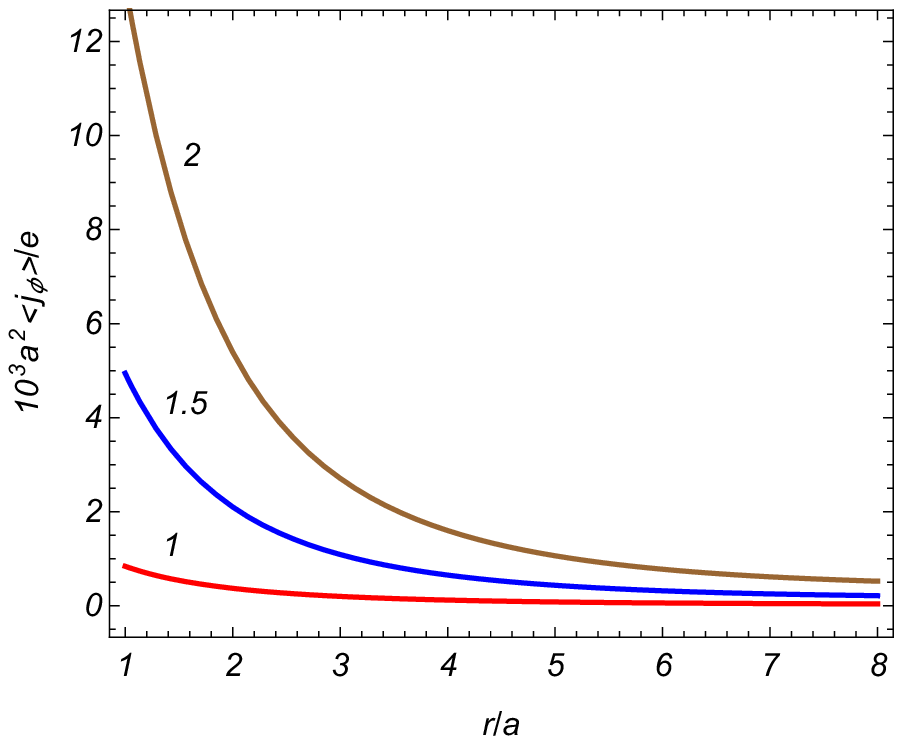,width=7.cm,height=5.5cm}%
\end{tabular}%
\end{center}
\caption{Charge and current densities versus $r/a$ for a massless fermionic
field. The graphs are plotted for $b/a=8$, $\protect\alpha _{0}=1/4$, $(s,%
\protect\lambda _{a},\protect\lambda _{b})=(1,1,1)$ and for different values
of the parameter $q$ (numbers near the curves). }
\label{fig4}
\end{figure}

Now we turn to the investigation of the radial dependence for the VEVs. In
figure \ref{fig4} the charge (left panel) and current (right panel)
densities are depicted as functions of $r/a$ for a massless field and and
boundary conditions with $\lambda _{a}=\lambda _{b}=1$. The graphs are
plotted for $b/a=8$, $\alpha _{0}=1/4$, and the numbers near the curves
correspond to the values of the parameter $q$. The curve for $q=1$
corresponds to a planar ring. As seen, the presence of the angle deficit may
essentially increase both the charge and current densities. For the example
presented in figure \ref{fig4} the ratio $\langle j^{0}\rangle /e$ is
negative near the inner edge and positive near the outer edge. The ratio $%
\langle j_{\phi }\rangle /e$ is positive.

\begin{figure}[tbph]
\begin{center}
\begin{tabular}{cc}
\epsfig{figure=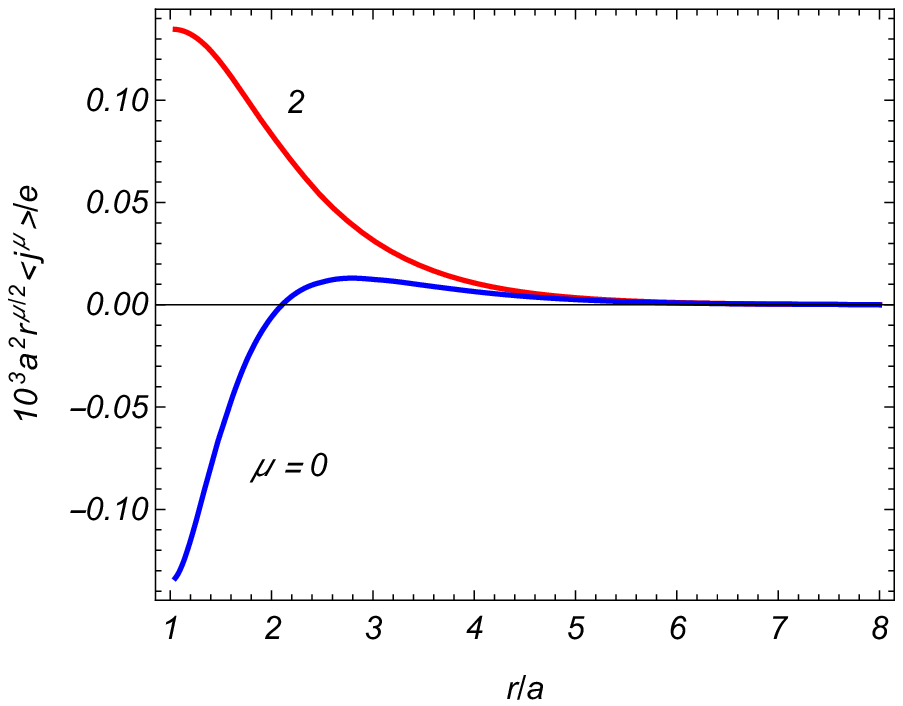,width=7.cm,height=5.5cm} & \quad %
\epsfig{figure=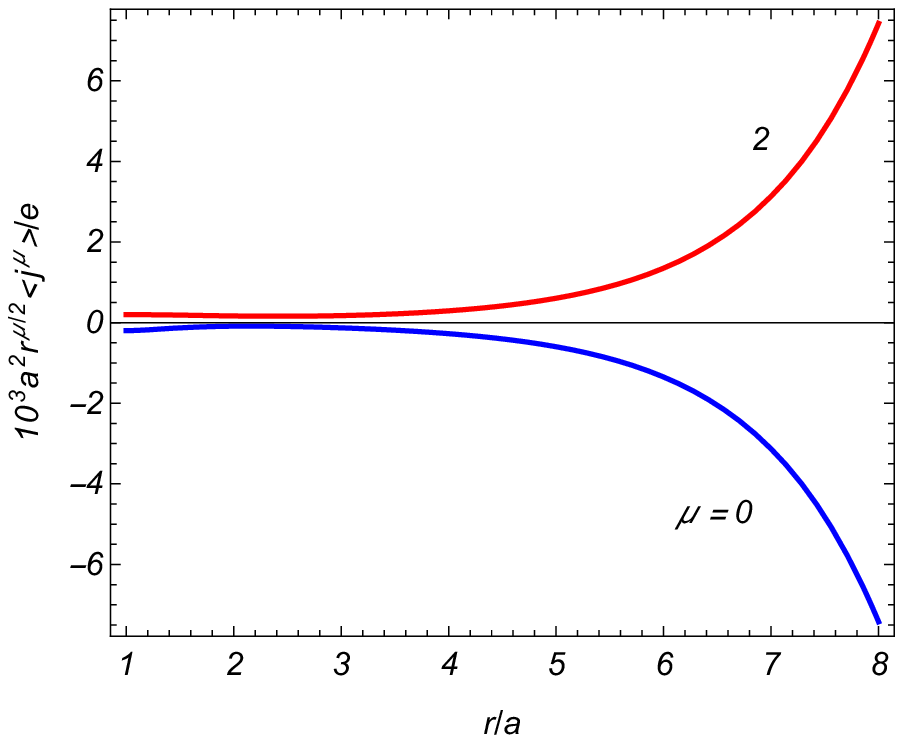,width=7.cm,height=5.5cm}%
\end{tabular}%
\end{center}
\caption{Charge ($\protect\mu =0$) and current ($\protect\mu =2$) densities
as functions of the radial coordinate for a massive fermionic field. The
graphs are plotted for $ma=0.5$, $b/a=8$, $\protect\alpha _{0}=1/4$. The
left and right panels correspond to the sets $(s,\protect\lambda _{a},%
\protect\lambda _{b})=(1,1,1)$ and $(s,\protect\lambda _{a},\protect\lambda %
_{b})=(1,1,-1)$, respectively. }
\label{fig5}
\end{figure}

\begin{figure}[tbph]
\begin{center}
\begin{tabular}{cc}
\epsfig{figure=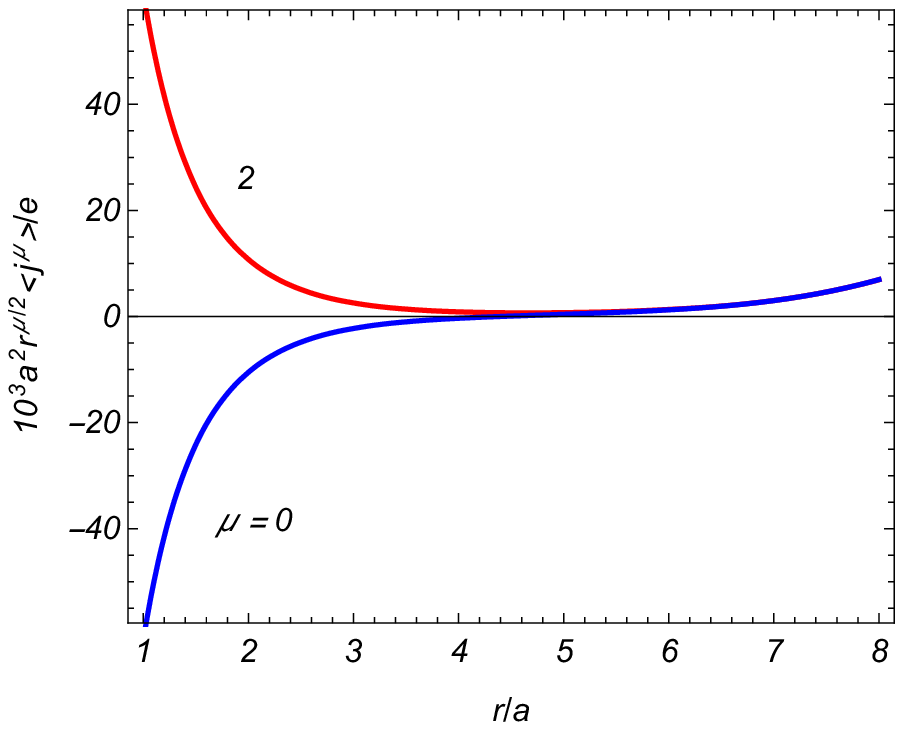,width=7.cm,height=5.5cm} & \quad %
\epsfig{figure=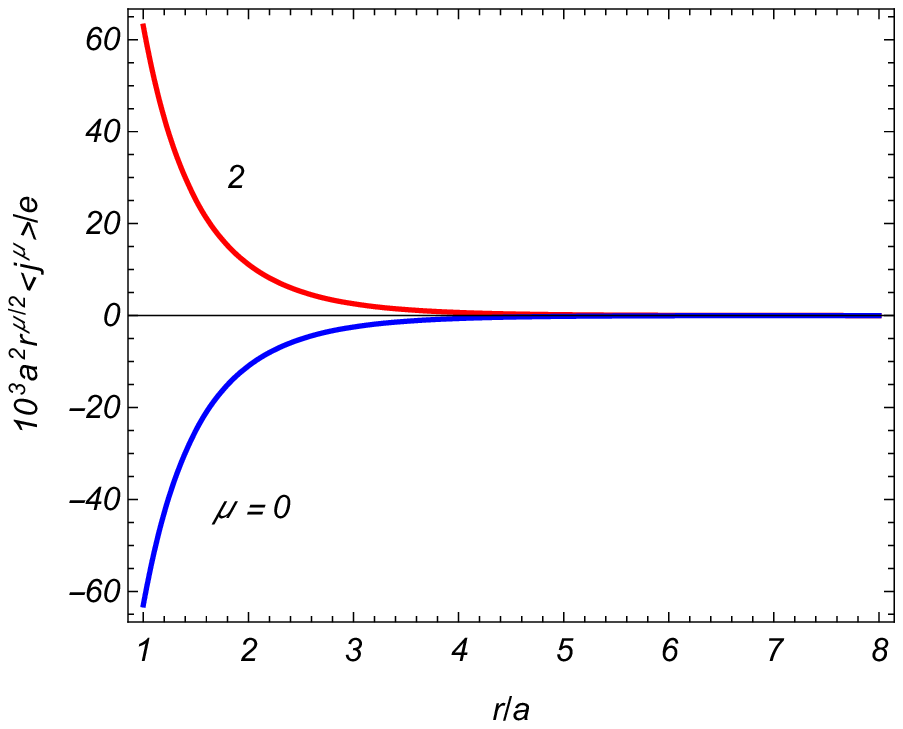,width=7.cm,height=5.5cm}%
\end{tabular}%
\end{center}
\caption{The same as in figure \protect\ref{fig5} for the sets $(s,\protect%
\lambda _{a},\protect\lambda _{b})=(-1,1,1)$ (left panel) and $(s,\protect%
\lambda _{a},\protect\lambda _{b})=(-1,1,-1)$ (right panel).}
\label{fig6}
\end{figure}

It is of interest to investigate the dependence of the VEVs on the values of
the parameters $(s,\lambda _{a},\lambda _{b})$. Figures \ref{fig5} and \ref%
{fig6} display the radial dependence of the charge and current densities for
different sets $(s,\lambda _{a},\lambda _{b})$ in the case of a massive
field with the mass corresponding to $ma=0.5$. The graphs are plotted for $%
q=1.5$, $b/a=8$, $\alpha _{0}=1/4$. The curves with $\mu =0$ correspond to
the charge density $\langle j^{0}\rangle $ and the curves $\mu =2$
correspond to the physical azimuthal component $\langle j_{\phi }\rangle
=r\langle j^{2}\rangle $ of the current density. Figure \ref{fig5}
corresponds to fields with $(s,\lambda _{a},\lambda _{b})=(1,1,1)$ (left
panel) and $(s,\lambda _{a},\lambda _{b})=(1,1,-1)$ (right panel). In figure %
\ref{fig6}, $(s,\lambda _{a},\lambda _{b})=(-1,1,1)$ for the left panel and $%
(s,\lambda _{a},\lambda _{b})=(-1,1,-1)$ for the right panel. The graphs for
other sets of the parameters $(s,\lambda _{a},\lambda _{b})$ are obtained
from the ones depicted in figures \ref{fig5} and \ref{fig6} by taking into
account that under the reflection $(s,\lambda _{a},\lambda _{b})\rightarrow
(-s,-\lambda _{a},-\lambda _{b})$ the charge density is an odd function and
the current density is an even function. As seen, the charge and current
densities are mainly located near the edges, inner or outer. The numerical
data confirm the relations (\ref{j2u}) between the charge and current
densities on the edges of the ring. An important point to mention here is
that the VEVs of the charge and current densities are finite on the edges of
ring. This is not the case, for example, for the fermion condensate or for
the VEV\ of the energy-momentum tensor.

\begin{figure}[tbph]
\begin{center}
\begin{tabular}{cc}
\epsfig{figure=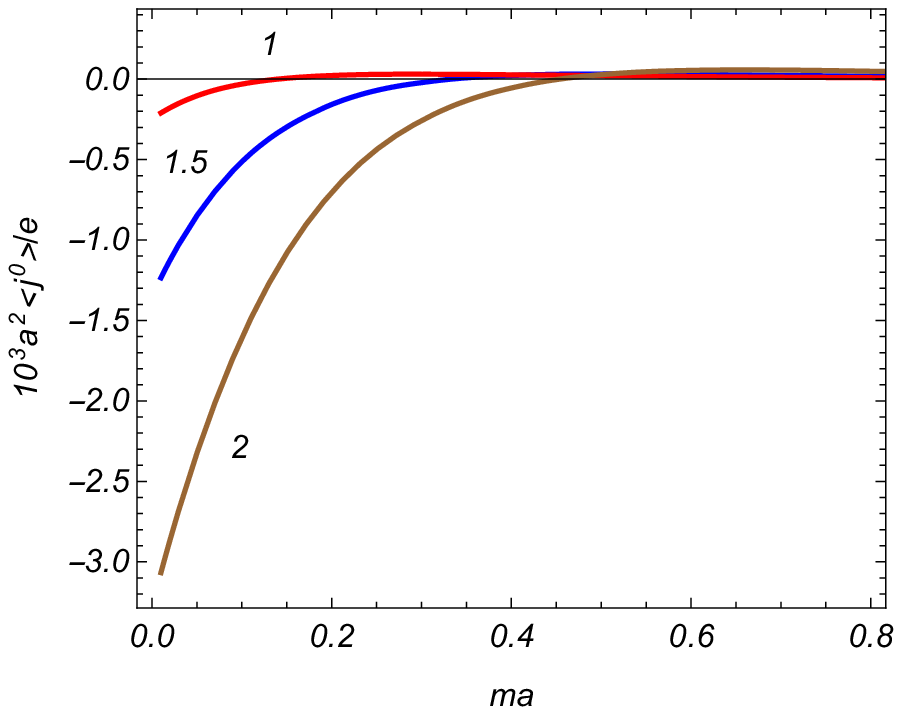,width=7.cm,height=5.5cm} & \quad %
\epsfig{figure=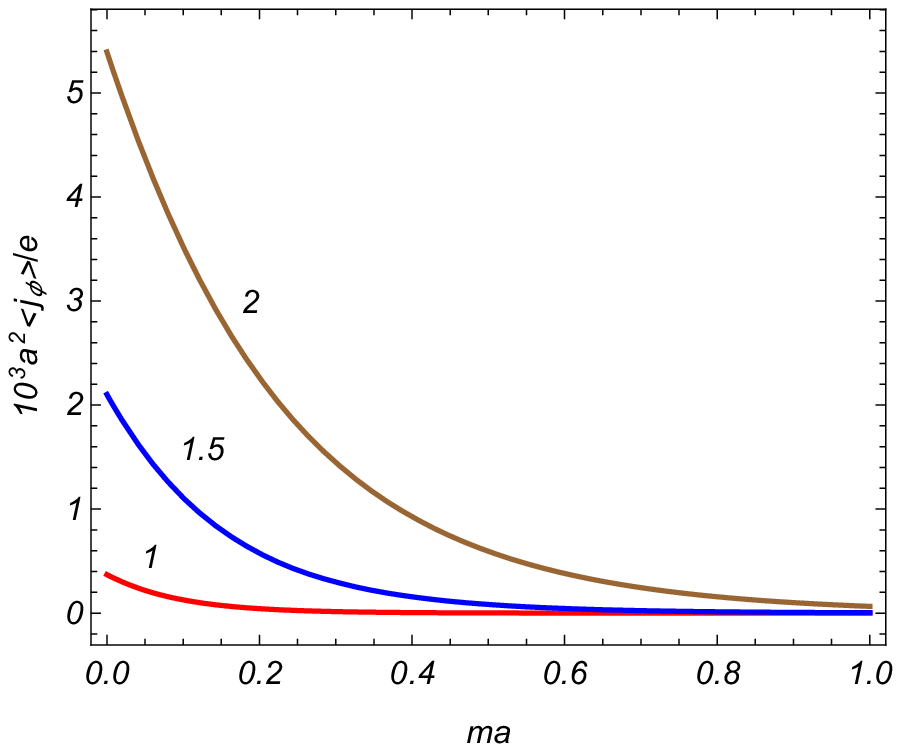,width=7.cm,height=5.5cm}%
\end{tabular}%
\end{center}
\caption{Charge and current densities as functions of the mass for $b/a=8$, $%
\protect\alpha _{0}=1/4$, $r/a=2$, $(s,\protect\lambda _{a},\protect\lambda %
_{b})=(1,1,1)$. The numbers near the curves are the corresponding values of $%
q$. }
\label{fig7}
\end{figure}

\begin{figure}[tbph]
\begin{center}
\begin{tabular}{cc}
\epsfig{figure=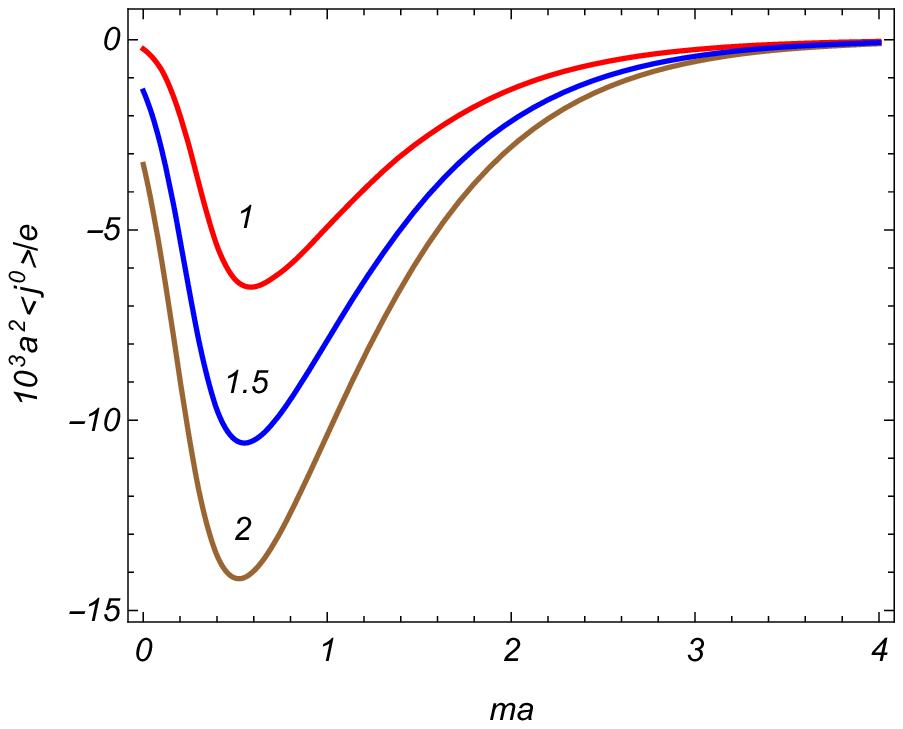,width=7.cm,height=5.5cm} & \quad %
\epsfig{figure=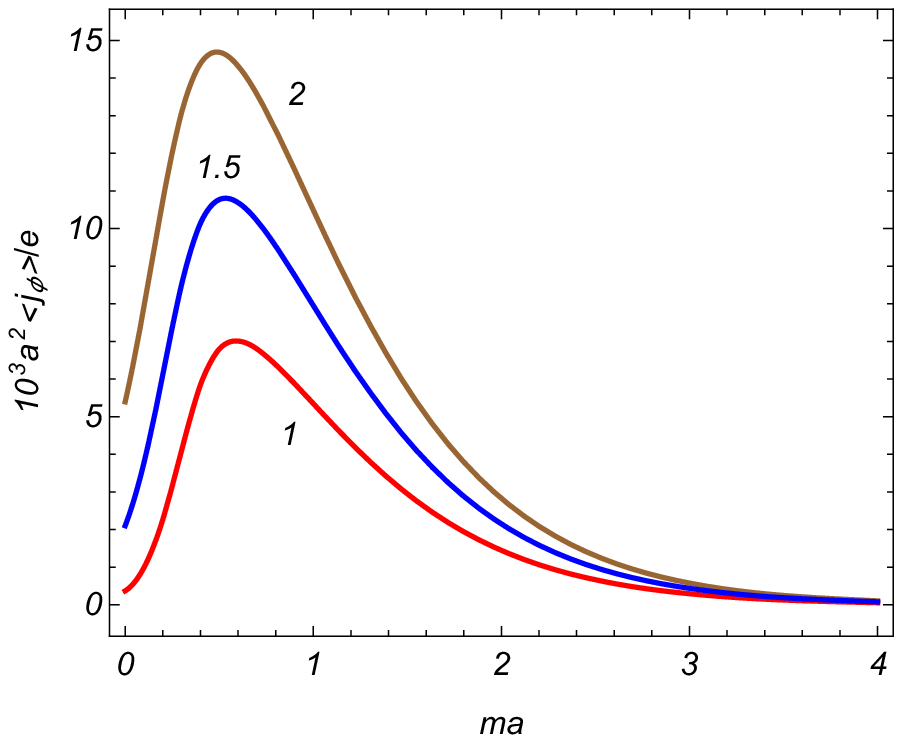,width=7.cm,height=5.5cm}%
\end{tabular}%
\end{center}
\caption{The same as in figure \protect\ref{fig7} for the field with $s=-1$.
}
\label{fig8}
\end{figure}

Comparing the left panel in figure \ref{fig5} with the graphs in figure \ref%
{fig4}, we see that for a massive field the VEVs are essentially smaller.
This can be not the case for other sets of the parameters $(s,\lambda
_{a},\lambda _{b})$. In order to see the dependence of the VEVs on the field
mass, in figure \ref{fig7} we plot the charge and current densities as
functions of $ma$ for fixed values $b/a=8$, $r/a=2$, $\alpha _{0}=1/4$ and
for the field with $s=1$. The numbers near the curves are the values for $q$%
. The same graphs for the field with $s=-1$ are presented in figure \ref%
{fig8}. As the numerical results show, the dependence on the mass is
essentially different for the cases $s=1$ and $s=-1$. For the parameters
corresponding to \ref{fig7} the VEVs decrease (by modulus) with increasing
mass. For the example corresponding to figure \ref{fig8} both the charge and
current densities increase by modulus with initial increase of the mass and
take their maximal or minimal values for some intermediate value of $ma$.
The further increase of the mass, as expected, leads to the suppression of
the VEVs.

\section{VEVs in parity and time-reversal invariant models and applications
to graphitic cones}

\label{sec:Tsym}

\subsection{VEVs for two irreducible representations of the Clifford algebra}

The fermionic field we have considered lives in two-dimensional space. In
even number of spatial dimensions there are two inequivalent irreducible
representations of the Clifford algebra. In this section it will be shown
how the VEVs of the charge and current densities are obtained from the
results given above for the fields realizing those representations. We will
distinguish the different representations by the parameter $s$ taking the
values $-1$ and $+1$ (as it will be seen below it coincides with the
parameter $s$ we introduced before in front of the mass term in the Dirac
equation (\ref{Dirac})). The corresponding sets of the $2\times 2$ Dirac
matrices will be denoted by $\gamma _{(s)}^{\mu }=(\gamma ^{0},\gamma
^{1},\gamma _{(s)}^{2})$ and the related fields by $\psi _{(s)}(x)$. In the
geometry described by the line element (\ref{linel}) the two inequivalent
representations of the matrix $\gamma _{(s)}^{2}$ can be taken as $\gamma
_{(s)}^{2}=-is\gamma ^{0}\gamma ^{1}/r$. For $s=+1$ the set $\gamma
_{(s)}^{\mu }$ coincides with (\ref{gamma}) used in the calculations above, $%
\gamma _{(+1)}^{\mu }=\gamma ^{\mu }$. For the Lagrangian density
corresponding to the fields $\psi _{(s)}(x)$ one has $L_{s}=\bar{\psi}%
_{(s)}(i\gamma _{(s)}^{\mu }D_{\mu }-m_{(s)})\psi _{(s)}$, where the mass
for different representations, in general, can be different. The current
densities of the fields are given by the standard formula $j_{(s)}^{\mu }=e%
\bar{\psi}_{(s)}\gamma _{(s)}^{\mu }\psi _{(s)}$. The boundary conditions on
the edges we will take again in the form (\ref{BCs}):
\begin{equation}
\left( 1+i\lambda _{r}^{(s)}n_{\mu }\gamma _{(s)}^{\mu }\right) \psi
_{(s)}(x)=0,\;r=a,b.  \label{BC}
\end{equation}%
Here, the parameters $\lambda _{r}^{(s)}$ also can be different for separate
representations.

Comparing with the discussion above, we see that for $\lambda
_{r}^{(+1)}=\lambda _{r}$ the field equation and the boundary conditions for
the field $\psi _{(+1)}(x)$ are the same as those for the field $\psi (x)$
with $s=1$ and $m=m_{(+1)}$, discussed in the previous sections. Hence, the
expressions of the VEVs of the charge and current densities for $\psi
_{(+1)}(x)$ coincide with those given above. In order to find the VEVs for
the field $\psi _{(-1)}(x)$, we introduce a new field $\psi _{(-1)}^{\prime
} $ according to $\psi _{(-1)}^{\prime }=\gamma ^{0}\gamma ^{1}\psi _{(-1)}$
with the inverse transformation $\psi _{(-1)}=\gamma ^{0}\gamma ^{1}\psi
_{(-1)}^{\prime }$. The Lagrangian density is written as $L_{-1}=\bar{\psi}%
_{(-1)}^{\prime }(i\gamma ^{\mu }D_{\mu }+m_{(-1)})\psi _{(-1)}^{\prime }$
with the gamma matrices $\gamma ^{\mu }=\gamma _{(+1)}^{\mu }$ and for the
current density we get $j_{(-1)}^{\mu }=e\bar{\psi}_{(-1)}^{\prime }\gamma
^{\mu }\psi _{(-1)}^{\prime }$. As seen, in terms of the new field the mass
term in the Lagrangian density reversed the sign. Substituting $\psi
_{(-1)}=\gamma ^{0}\gamma ^{1}\psi _{(-1)}^{\prime }$ in the boundary
condition (\ref{BC}) for $s=-1$, we get the corresponding condition for the
primed field $\psi _{(-1)}^{\prime }(x)$:%
\begin{equation}
\left( 1-i\lambda _{r}^{(-1)}n_{\mu }\gamma ^{\mu }\right) \psi
_{(-1)}^{\prime }=0,  \label{BCm}
\end{equation}%
for $r=a,b$. From these considerations it follows that the charge and
current densities for the field $\psi _{(-1)}(x)$ are obtained from the
expressions given above taking $s=-1$ and $\lambda _{r}=-\lambda _{r}^{(-1)}$%
.

\subsection{Charge and current densities in parity and time-reversal
symmetric models}

In two spatial dimensions, the mass term in the Lagrangian density for a
two-component fermionic field $\psi (x)$ is not invariant under the parity ($%
P$) and time-reversal ($T$) transformations. In the absence of magnetic
fields, $P$- and $T$-symmetric models can be constructed combining two
fields realizing different irreducible representations of the Clifford
algebra and having the same mass. In accordance with the consideration of
the previous subsection, the Lagrangian density for this set of fields,
denoted as before by $\psi _{(s)}$, $s=\pm 1$, is written in two equivalent
forms%
\begin{eqnarray}
L &=&\sum_{s=\pm 1}\bar{\psi}_{(s)}(i\gamma _{(s)}^{\mu }D_{\mu }-m)\psi
_{(s)}  \notag \\
&=&\sum_{s=\pm 1}\bar{\psi}_{(s)}^{\prime }(i\gamma ^{\mu }D_{\mu }-sm)\psi
_{(s)}^{\prime },  \label{Lag2}
\end{eqnarray}%
where $\psi _{(+1)}^{\prime }=\psi _{(+1)}$ and $\psi _{(-1)}^{\prime
}=\gamma ^{0}\gamma ^{1}\psi _{(-1)}^{\prime }$. The total current density
is given by the formula $J^{\mu }=e\sum_{s=\pm 1}\bar{\psi}_{(s)}\gamma
_{(s)}^{\mu }\psi _{(s)}$ or by $J^{\mu }=e\sum_{s=\pm 1}\bar{\psi}%
_{(s)}^{\prime }\gamma ^{\mu }\psi _{(s)}^{\prime }$. The separate fields
obey the boundary conditions (\ref{BC}) or the conditions $\left(
1+is\lambda _{r}^{(s)}n_{\mu }\gamma ^{\mu }\right) \psi _{(s)}^{\prime
}(x)=0$ in terms of the primed fields. Note that because of the appearance
of the factor $s$ in front of the term with the normal to the boundary, the
fields $\psi _{(+1)}^{\prime }$ and $\psi _{(-1)}^{\prime }$ obey different
boundary conditions if the fields $\psi _{(+1)}$ and $\psi _{(-1)}$ are
constrained by the same boundary conditions and vice versa.

We can combine the two-component fields $\psi _{(s)}(x)$ in a single
4-component spinor field $\Psi =(\psi _{(+1)},\psi _{(-1)})^{T}$ with the
Lagrangian density
\begin{equation}
L=\bar{\Psi}(i\gamma _{(4)}^{\mu }D_{\mu }-m)\Psi ,  \label{Lag3}
\end{equation}%
where the $4\times 4$ Dirac matrices are given by $\gamma _{(4)}^{\mu
}=I\otimes \gamma ^{\mu }$ for $\mu =0,1$, and $\gamma _{(4)}^{2}=\sigma
_{3}\otimes \gamma ^{2}$ with $\sigma _{3}$ being the Pauli matrix. For the
corresponding current density one has the standard expression $J^{\mu }=e%
\bar{\Psi}(x)\gamma _{(4)}^{\mu }\Psi (x)$. The boundary conditions on the
edges $r=a,b$ are rewritten as
\begin{equation}
\left( 1+i\Lambda _{r}n_{\mu }\gamma _{(4)}^{\mu }\right) \Psi (x)=0,
\label{BCPsi}
\end{equation}%
with $\Lambda _{r}=\mathrm{diag}(\lambda _{r}^{(+1)},\lambda _{r}^{(-1)})$.
Alternatively, we can introduce the spinor $\Psi ^{\prime }=(\psi
_{(+1)}^{\prime },\psi _{(-1)}^{\prime })^{T}$ and the set of gamma matrices
$\gamma _{(4)}^{\prime \mu }=\sigma _{3}\otimes \gamma ^{\mu }$. For the
corresponding Lagrangian density one gets $L=\bar{\Psi}^{\prime }(i\gamma
_{(4)}^{\prime \mu }D_{\mu }-m)\Psi ^{\prime }$ and for the current density
operator $J^{\mu }=e\bar{\Psi}^{\prime }(x)\gamma _{(4)}^{\mu }\Psi ^{\prime
}(x)$. Now the boundary conditions take the form $\left( 1+i\Lambda
_{r}n_{\mu }\gamma _{(4)}^{\prime \mu }\right) \Psi ^{\prime }(x)=0$. The
latter has the same form as (\ref{BCPsi}), though with different
representation of the gamma matrices.

For $\lambda _{u}^{(+1)}=\lambda _{u}^{(-1)}$, $u=a,b$, the fields $\psi
_{(+1)}$ and $\psi _{(-1)}$ in the Lagrangian density (\ref{Lag2}) obey the
same boundary conditions. In this case the boundary condition for the
transformed field $\psi _{(-1)}^{\prime }$ differs from the condition for
the field $\psi _{(+1)}^{\prime }=\psi _{(+1)}$ by the sign of the term
containing the normal to the boundary. As it has been shown above, the
charge density is an odd function under the replacement $(s,\lambda
_{u})\rightarrow (-s,-\lambda _{u})$, whereas the azimuthal current density
is an even function. From here we conclude that in the model involving two
fields $\psi _{(+1)}$ and $\psi _{(-1)}$ with the same masses and the phases
in the periodicity condition (\ref{PC}), obeying the boundary conditions (%
\ref{BC}) with $\lambda _{u}^{(+1)}=\lambda _{u}^{(-1)}$, the VEV of the
total charge density vanishes, $\langle J^{0}\rangle =0$, and for the VEV of
the total current density one gets $\langle J^{2}\rangle =2\langle
j^{2}\rangle $, where $\langle j^{2}\rangle $ is given by (\ref{jmu3}) with $%
\mu =2$ and with $s=1$, $\lambda _{u}=\lambda _{u}^{(+1)}$.

In models with two fields $\psi _{(+1)}$ and $\psi _{(-1)}$, realizing
inequivalent irreducible representations of the Clifford algebra, a nonzero
vacuum charge density may appear if the corresponding boundary conditions
are different ($\lambda _{u}^{(+1)}\neq \lambda _{u}^{(-1)}$) or the masses
for the fields differ. However, note that the difference in the masses will
break the parity and time-reversal symmetry of the model. Another
possibility for the appearance of the nonzero charge density is realized in
models with different phases in the periodicity conditions (\ref{PC}) for
the fields $\psi _{(+1)}$ and $\psi _{(-1)}$. The latter type of situation
takes place in semiconducting carbon nanotubes where the fields under
consideration describe the electronic subsystem of graphene tubes.

\subsection{Current density in graphitic cones}

Among important realizations of 2D fermionic models is graphene. The
existence of various classes of graphene allotropes, like carbon nanotubes,
fullerens, graphitic cones, nanoloops and nanohorns, makes graphene an
exciting arena for the investigation of the effects of the geometry,
topology and boundaries on the properties of a quantum fermionic field.
Recently, a number of mechanisms have been suggested (see, for example, \cite%
{Volo15}) to generate effective curved background geometries for Dirac
fermions in graphene. In particular, they include various types of external
fields, lattice deformations, and local variations of the Fermi velocity.
The advantage of these graphene based artificial systems in modelling the
influence of the gravity on quantum matter is that one can tune in a
controlled manner the geometrical characteristics of the background
spacetime.

In the long wavelength approximation, the effective field theory for the
electronic subsystem in graphene is formulated in terms of 4-component
spinors $\Psi _{S}=(\psi _{+,AS},\psi _{+,BS},\psi _{-,AS},\psi _{-,BS})^{T}$%
, where $S=\pm 1$ corresponds to the spin degree of freedom. It is
decomposed into two 2-component spinors, $\psi _{+}=(\psi _{+,AS},\psi
_{+,BS})$ and $\psi _{-}=(\psi _{-,AS},\psi _{-,BS})$, corresponding to two
inequivalent corner points $\mathbf{K}_{+}$ and $\mathbf{K}_{-}$ of the
hexagonal Brillouin zone of graphene. These two valleys are related by the
time-reversal symmetry. The separate components $\psi _{\pm ,AS}$ and $\psi
_{\pm ,BS}$ give the amplitude of the electron wave function on the
triangular sublattices $A$ and $B$ of the graphene hexagonal lattice. In the
standard units with the speed of light $c$ and the Planck constant $\hbar $,
the Lagrangian density in the effective field theory is presented as
\begin{equation}
L_{\mathrm{g}}=\sum_{S=\pm 1}\bar{\Psi}_{S}[i\hbar \gamma _{(4)}^{0}\partial
_{t}+i\hbar v_{F}\gamma _{(4)}^{l}(\nabla _{l}+ieA_{l}/\hbar c)-\Delta ]\Psi
_{S},  \label{Lgraph}
\end{equation}%
where $l=1,2$, $e$ is the electron charge and $v_{F}\approx 7.9\times 10^{7}$
cm/s is the Fermi velocity for electrons. The energy gap $\Delta $,
introduced in (\ref{Lgraph}), is related to the Dirac mass $m$ by $\Delta
=mv_{F}^{2}$. A number of mechanisms has been considered in the literature
for the generation of the energy gap in the range $1\,\mathrm{meV}\lesssim
\Delta \lesssim 1\,\mathrm{eV}$ (see, for example, \cite{Gusy07} and
references therein). The energy scale in the model is determined by the
parameter $\gamma _{F}=$ $\hbar v_{F}/a_{0}\approx 2.51\,\mathrm{eV}$, where
$a_{0}\approx 1.42$ \AA\ is the inter-atomic spacing of graphene honeycomb
lattice. For the Compton wavelength related to the energy gap one has $%
a_{C}=\hbar v_{F}/\Delta $. For a given $S$, the charge density
corresponding to the Lagrangian (\ref{Lgraph}) is given by $J^{0}=e\bar{\Psi}%
_{S}(x)\gamma _{(4)}^{0}\Psi _{S}(x)$ and for the current density we get $%
J^{\mu }=ev_{F}\bar{\Psi}_{S}(x)\gamma _{(4)}^{\mu }\Psi _{S}(x)$, $\mu =1,2$%
.

The separate parts in (\ref{Lgraph}) for given $S$ are the analog of the
Lagrangian density (\ref{Lag3}) we have discussed before. The two-component
fields $\psi _{(+1)}$ and $\psi _{(-1)}$ correspond to the fields $\psi _{+}$
and $\psi _{-}$. Hence, the parameter $s$ in the discussion above enumerates
the valley degrees of freedom in graphene. On the base of this analogy, we
can apply the formulas for the charge and current densities given above to
graphene conical ribbons with the edges $r=a$ and $r=b$. The graphene
nanocones have attracted considerable attention due to their potential
applications such as probes for scanning probe microscopy, electron
emitters, tweezers for nanomanipulation, energy storage, gas sensors, and
biosensors. In the problem under consideration the separate parts with $%
S=\pm 1$ give the same contributions to the VEVs and we can consider the
VEVs for a given spin degree of freedom omitting the index $S$. The total
VEVs are obtained with an additional factor 2. As it has been already
mentioned in Introduction, for the opening angle in graphitic cones one has $%
\phi _{0}=2\pi (1-n_{c}/6)$ with $n_{c}=1,2,\ldots ,5$ being the number of
the removed sectors from a planar graphene sheet. The analog of the
quasiperiodicity condition (\ref{PC}) in graphene cones has been discussed
in \cite{Lamm00,Lamm04,Site07,Chak11}. For graphene cones with odd values of
$n_{c}$ it mixes the valley indices through the factor $e^{-i\pi n_{c}\tau
_{2}/2}$, where the Pauli matrix $\tau _{2}$ acts on those indices. The
corresponding condition can be diagonalized by a unitary transformation that
diagonalizes the matrix $\tau _{2}$. For even values of $n_{c}$ the spinors
corresponding to different valleys are not entwined and an additional
diagonalization is not required. By taking into account that only the
fractional part of the parameter $\chi $ is relevant in the evaluation of
the VEVs, it can be seen that two inequivalent values of the parameter $\chi
$ realized in graphitic cones correspond to $\chi =\pm 1/3$. Note that the
same inequivalent values of the periodicity phase are realized in
semiconducting carbon nanotubes (in metallic nanotubes $\chi =0$). The
fermionic current density in cylindrical and toroidal carbon tubes has been
investigated in \cite{Bell10,Bell13}.

For a given spin $S$, the ground state charge and current densities in
graphitic cones are obtained from the results in section \ref{sec:jmu} in
accordance of the procedure described in the previous subsection, adding an
additional factor $v_{F}$ for the azimuthal current density. Translating the
results given above to graphene made structures it is convenient to make the
replacements $mu\rightarrow u/a_{C}$, $u=a,b,r$, in the corresponding
formulas. If the energy gap is the same for both the valleys, the net charge
density vanishes as a consequence of the cancellation between the
contributions from different valleys. However, there exist gap generations
mechanisms in graphene breaking the valley symmetry (for example, chemical
doping) and one can have a situation with different masses for the fields $%
\psi _{+}$ and $\psi _{-}$. In this case there is no cancellation of the
corresponding contributions to the charge density. Note that the magnetic
flux induced currents in planar graphene rings have been investigated in
\cite{Bell16,Rech07}. Based on the concept of branes, a model for the
emergence of current density in graphene in the presence of defects has been
recently discussed in \cite{Sepe17}.

\section{Conclusion}

\label{sec:Conc}

The notion of vacuum in quantum field theory has a global nature and its
properties are sensitive to both the local and global characteristics of the
background spacetime. In the present paper we have investigated the combined
effects of boundaries, topology and of the magnetic flux on the ground state
mean charge and current densities for a fermionic field in two-dimensional
conical rings with arbitrary values of the angle deficit. The boundary
conditions for the field operator on the ring edges are specified by the set
of parameters $(\lambda _{a},\lambda _{b})$. In the special case $(1,1)$
they are reduced to the standard MIT bag boundary condition (infinite mass
boundary condition in the context of 2D fermionic systems). An additional
parameter $s$ in front of the mass term in the Dirac equation corresponds to
two inequivalent irreducible representations of the Clifford algebra in
(2+1)-dimensional spacetime. The fermionic mode functions are presented as (%
\ref{psie}), where the allowed values of the radial quantum number depend on
the specific boundary condition and are roots of equation (\ref{EigEq}). For
whole family of boundary conditions, we have considered, the vacuum state is
stable and for all the roots $\gamma ^{2}\geq -m^{2}$. For fields with $%
(s,\lambda _{a},\lambda _{b})=\left( \pm 1,\pm 1,\pm 1\right) $ all the
eigenvalues for $\gamma $ are real. In the remaining cases, depending on $%
b/a $ and $ma$, purely imaginary eigenvalues $\gamma =i\eta /a$, $0<\eta <ma$%
, may appear corresponding to bound states. For half-integer values of the
parameter $\alpha $ from (\ref{alpha}) and under the condition $\lambda
_{a}=-\lambda _{b}$ there is also a zero mode with the value of the total
angular momentum $j=-\alpha $.

The VEVs of the charge and current densities are evaluated by using the
corresponding mode sums over the bilinear products of the mode functions.
The VEV for the radial current vanishes and the contribution of the modes
with positive $\gamma $ to the charge and azimuthal current densities is
presented as (\ref{jmu}). In the presence of the bound state or the zero
mode, the corresponding contributions, given by (\ref{jb}) and (\ref{j0s10}%
), should be added to (\ref{jmu}). The charge and current densities on the
ring edges are connected by simple relation (\ref{j2u}) that is valid for
whole family of boundary conditions. For half-integer values of $\alpha $
the charge and current densities vanish for the boundary conditions with $%
\lambda _{a}=\lambda _{b}$. For the conditions with $\lambda _{a}=-\lambda
_{b}$ the only nonzero contribution comes from the zero mode. In the latter
case the charge and current densities are discontinuous functions of $\alpha
$ (in particular, of the magnetic flux enclosed by the ring) at half-integer
values of that parameter.

In the representation (\ref{jmu}) the summation goes over the eigenvalues
for $\gamma $ given implicitly, as roots of equation (\ref{EigEq}). The
explicit knowledge of those roots is not required if we apply the summation
formula (\ref{Sum}) to the corresponding series. In the presence of the
bound states an additional term in the form (\ref{BScontr2}) should be added
to the right-hand side of (\ref{Sum}). We have shown that the additional
term exactly cancels the contribution coming from the bound states and the
integral representation (\ref{jmu2}) is valid for all the sets of parameters
$(s,\lambda _{a},\lambda _{b})$. The first term in the right-hand side of (%
\ref{jmu2}) corresponds to the VEV in the conical geometry with a single
boundary at $r=a$ and the last term is interpreted as the contribution
induced by the second edge at $r=b$. The former part is further decomposed
as (\ref{j0adec}) with the boundary-free and edge induced contributions,
given by (\ref{jm02}) and (\ref{jmua2}), respectively. An alternative
representation, where the part corresponding to the problem inside a single
circular boundary is extracted, is given by (\ref{jmu4}). As a general rule,
the modulus of both the charge and current densities increases with
increasing planar angle deficit (with increasing $q$). Depending on the
boundary condition, determined by the set $(\lambda _{a},\lambda _{b})$, the
charge and current densities are mainly located near the inner or outer edge
(see figures \ref{fig4}-\ref{fig6}). We have demonstrated that the behavior
of the VEVs as functions of the mass can be essentially different for fields
with $s=+1$ and $s=-1$. In the former case and for the boundary condition
with $(\lambda _{a},\lambda _{b})=(1,1)$ the absolute values of the charge
and current densities decrease with increase of the field mass. In the case $%
s=-1$ and for the same boundary condition, the absolute values for both the
charge and current densities increase with initial increase of the mass.
After taking the maximum value, as expected, they tend to zero for large
masses.

It is well known that in two spatial dimensions the fermionic mass term
breaks both the parity and time reversal invariances. $P$- and $T$-symmetric
fermionic models are constructed considering the set of two fields, $\psi
_{(+1)}$ and $\psi _{(-1)}$, with the same masses realizing two inequivalent
irreducible representations of the Clifford algebra. The VEVs of the charge
and current densities for the field corresponding to the second
representation and obeying the boundary condition (\ref{BC}) are obtained
from the formulas in section \ref{jmu} with $s=-1$ and $\lambda
_{u}=-\lambda _{u}^{(-1)}$, $u=a,b$. If in addition to the masses, the
phases in the periodicity condition along the azimuthal direction and the
boundary conditions on the edges for the fields $\psi _{(+1)}$ and $\psi
_{(-1)}$ are the same then the total charge density vanishes, whereas the
total current density doubles. In the effective low-energy theory for
electronic subsystem of graphene, the fields $\psi _{(+1)}$ and $\psi
_{(-1)} $ correspond to two inequivalent points of the Brillouin zone
(valley degrees of freedom) and the results obtained in the present paper
can be applied for the investigation of the charge and current densities
induced by Aharonov-Bohm magnetic flux in graphitic cones. Two inequivalent
values of the phase $2\pi \chi $ realized in graphitic cones correspond to $%
\pm 2\pi /3 $ and for the parameter $q$ one has $q=1/(1-n_{c}/6)$. It is of
interest to note that the valley-dependent gap generation mechanisms (for a
recent discussion see \cite{Lu16} and references therein) create different
masses for the fields $\psi _{(+1)}$ and $\psi _{(-1)}$ and, as a result of
that, the nonzero net charge density appears. This breaks the time-reversal
symmetry.

\section*{Acknowledgments}

A.A.S. was supported by Viktor Ambartsumian Research Fellowship 2019-2020.
I.B. and A.A.S. are grateful for support of this project by The Norwegian
Research Council (Project 250346). A.A.S. gratefully acknowledges the
hospitality of the INFN, Laboratori Nazionali di Frascati (Frascati, Italy)
and the Norwegian University of Science and Technology (Trondheim, Norway),
where a part of this work was done. H.G.S. was supported by the grant No.
18T-1C355 of the Committee of Science of the Ministry of Education and
Science RA.

\appendix

\section{Contribution of the bound states}

\label{sec:App1}

In addition to the infinite set of positive modes $\gamma =\gamma _{l}$,
depending on the parameters of the model, the equation (\ref{EigEq}) for the
eigenmodes may have purely imaginary solutions $\gamma a=i\eta $, $\eta >0$.
For the modes with $\eta >ma$ the corresponding energy is imaginary and the
presence of these modes would signal about the instability of the vacuum
state. In the case $\eta >ma$ the equation determining the modes is given by%
\begin{equation}
K_{n_{p}}^{(a)}\left( \eta \right) I_{n_{p}}^{(b)}\left( \eta b/a\right)
-I_{n_{p}}^{(a)}(\eta )K_{n_{p}}^{(b)}(\eta b/a)=0.  \label{ImEq}
\end{equation}%
The left-hand side is a complex function and the real and imaginary parts
should be separately zero. Introducing the function%
\begin{equation}
B_{\mu ,\nu }(x,y)=I_{\mu }(x)K_{\nu }(y)-(-1)^{\mu -\nu }K_{\mu }(x)I_{\nu
}(y),  \label{Bmu}
\end{equation}%
from those conditions it follows that we should have
\begin{equation}
B_{n_{p}+1,n_{p}+1}\left( \eta ,\eta b/a\right) =\lambda _{a}\lambda
_{b}B_{n_{p},n_{p}}\left( \eta ,\eta b/a\right) .  \label{RelB}
\end{equation}%
By taking into account that $B_{\nu ,\nu }\left( \eta ,\eta b/a\right) <0$
for $\nu \geq -1/2$, we see that the equation (\ref{RelB}) has no solutions
for $\lambda _{a}\lambda _{b}<0$. For $\lambda _{a}\lambda _{b}>0$, noting
that $B_{\nu +1,\nu +1}\left( \eta ,\eta b/a\right) <B_{\nu ,\nu }\left(
\eta ,\eta b/a\right) $ for $\nu >-1/2$, again, (\ref{RelB}) has no
solutions. Hence, for all the values of the parameters $s,\lambda
_{a},\lambda _{b}$ there are no modes with $\eta >ma$ and the vacuum state
is stable.

Now we turn to the modes $\gamma a=i\eta $, $\eta >0$, with $\eta <ma$.
These modes correspond to bound states. They are determined by the equation (%
\ref{EigEq}) with $\gamma a=i\eta $. Introducing the modified Bessel
functions it is written in the form%
\begin{equation}
G_{\beta _{j}}(\eta ,\eta b/a)\equiv I_{\beta _{j}}^{(a)}\left( \eta \right)
K_{\beta _{j}}^{(b)}\left( \eta b/a\right) -I_{\beta _{j}}^{(b)}\left( \eta
b/a\right) K_{\beta _{j}}^{(a)}\left( \eta \right) =0,  \label{Gb}
\end{equation}%
where the functions $f_{\beta _{j}}^{(u)}(z)$ with $f=I,K$ are defined by (%
\ref{fnu}) with the replacement
\begin{equation}
i\sqrt{z^{2}-m_{u}^{2}}\rightarrow \kappa \sqrt{m_{u}^{2}-z^{2}}.
\label{replb}
\end{equation}%
This replacement is understood also in the following formulas in this
appendix.

If we write the roots of (\ref{Gb}) as functions of the parameters, $\eta
=\eta (b/a,s,\lambda _{u},j,\alpha _{0},\kappa )$, then the solutions for
different sets of the parameters in the arguments are connected by the same
relations as those for the modes $z_{l}$ (see the paragraph after formula (%
\ref{alpha})). For all values of the ratio $b/a$ the bound states are absent
in the cases $(s,\lambda _{a},\lambda _{b})=(\pm 1,\pm 1,\pm 1)$. For the
remaining sets, depending on the values of the parameters, we have the
following two possibilities: (i) the bound states are present for all values
of the ratio $b/a$ or (ii) they appear started from some critical value of
that parameter, denoted here as $(b/a)_{\mathrm{c}}$. The numerical analysis
has shown the following features. If there is no bound state for some $j=j_{(%
\mathrm{b})}$, then there is no bound state for angular quantum numbers with
$|j|>|j_{(\mathrm{b})}|$. The critical values of $b/a$ for the appearance of
the bound states increase with decreasing $ma$. The critical value $(b/a)_{%
\mathrm{c}}$ also increases with increasing $q$. The latter means that we
can have a situation when the bound state is present in a planar ring and is
absent in the conical ring for the same values of the other parameters. For
example, for $\kappa =+$, $ma=0.5$, $\alpha _{0}=1/4$ and $(s,\lambda
_{a},\lambda _{b})=(-1,1,1)$ for a planar ring ($q=1$) one has $(b/a)_{%
\mathrm{c}}\approx 3.13,4.5$ for $j=1/2,3/2$, respectively. For a conical
ring we get $(b/a)_{\mathrm{c}}\approx 3.6,6.22$ for $j=1/2,3/2$. For $%
(s,\lambda _{a},\lambda _{b})=(-1,1,1)$ and for the same values of the other
parameters the bound states are present only for $q=1$, $j=1/2$ with the
critical value $(b/a)_{\mathrm{c}}\approx 4.88$ and and there are no bound
states for $q=1.5$.

\begin{figure}[tbph]
\begin{center}
\begin{tabular}{cc}
\epsfig{figure=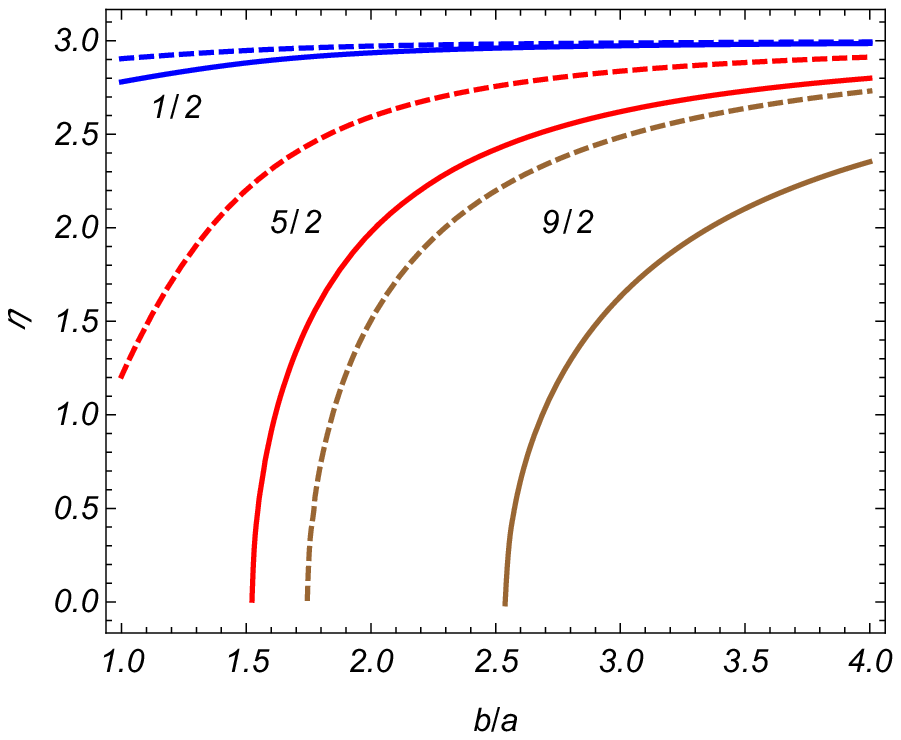,width=7.cm,height=5.5cm} & \quad %
\epsfig{figure=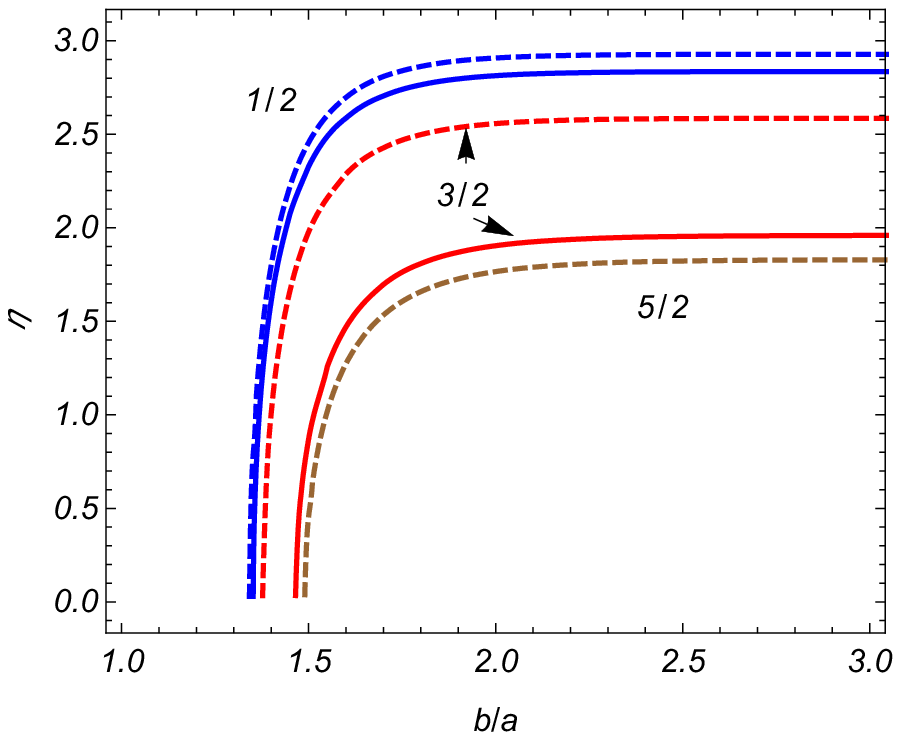,width=7.cm,height=5.5cm}%
\end{tabular}%
\end{center}
\caption{The values of $\protect\eta $ for bound states versus the ratio $%
b/a $. The numbers near the curves are the values of the quantum number $j$.
The left and right panels are plotted for the values of the parameters $(s,%
\protect\lambda _{a},\protect\lambda _{b})=(-1,1,1)$ and $(s,\protect\lambda %
_{a},\protect\lambda _{b})=(-1,-1,1)$, respectively. The full and dashed
curves correspond to the cases $q=1.5$ and $q=1$. For the other parameters
we have taken $\protect\alpha _{0}=1/4$ and $ma=3$. }
\label{fig9}
\end{figure}

In the limit $b/a\rightarrow \infty $ the equation for the bound states is
reduced to $K_{\beta _{j}}^{(a)}\left( \eta \right) =0$. The latter is the
equation for the bound states in a conical space with a single boundary at $%
r=a$ and has no solutions for $s\lambda _{a}>0$. In this case, in the limit $%
b/a\rightarrow \infty $ the possible bound states determined from (\ref{Gb})
tend to $ma$. If there is a bound state in the geometry of a single
boundary, then in the limit $b/a\rightarrow \infty $ the corresponding bound
state for a conical ring (with the same values for the set $(s,\lambda
_{a},j,\alpha _{0},\kappa )$) tends to the limiting value different from $ma$%
. These two situations are illustrated in figure \ref{fig9}, where we have
plotted the radial quantum number $\eta $ for the bound states as a function
of the ratio $b/a$ for $(s,\lambda _{a},\lambda _{b})=(-1,-1,1)$ (left
panel) and $(s,\lambda _{a},\lambda _{b})=(1,-1,-1)$ (right panel). The
graphs are plotted for $\kappa =+$, $ma=3$ and $\alpha _{0}=1/4$ and the
numbers near the curves are the values of $j$. The dashed and full curves
correspond to $q=1$ (planar ring) and $q=1.5$, respectively. For the left
panel $s\lambda _{a}>0$ and there is no bound state in a conical geometry
with a single boundary, $r\geq a$. In this case the bound states tend to $ma$
for $b/a\gg 1$. For the right panel the equation $K_{\beta _{j}}^{(a)}\left(
\eta \right) =0$ has a solution and it is the limiting value of the bound
state when $b/a\rightarrow \infty $. On the right panel we also see that the
bound states appear only started from some critical value of $b/a$. By using
the relations between the bound states for different sets of the parameters,
we see that the graphs in figure \ref{fig9} also present the locations of
the bound states for the set $(-j,-\alpha _{0},-\kappa )$ or for the set $%
(-s,-\lambda _{u},-\kappa )$ with the same values of the remaining
parameters.

\begin{figure}[tbph]
\begin{center}
\epsfig{figure=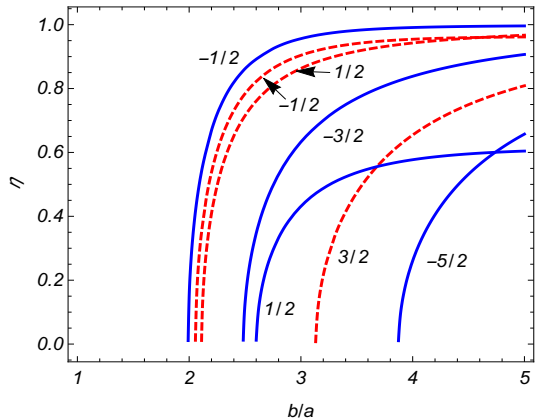,width=7.cm,height=5.5cm}
\end{center}
\caption{The values of $\protect\eta $ corresponding to the bound states
versus $b/a$ for $q=1.5$, $\protect\alpha _{0}=1/4$, $ma=1$, $(s,\protect%
\lambda _{a},\protect\lambda _{b})=(1,-1,-1)$. The full and dashed curves
correspond to positive and negative energy states and the numbers near the
curves are the values of $j$.}
\label{fig10}
\end{figure}

It is of interest to compare the number of the positive and negative energy
bound states for given values of the other parameters. In figure \ref{fig10}
the bound states are presented as functions of $b/a$ for $q=1.5$, $\alpha
_{0}=1/4$, $ma=1$, and for the set $(s,\lambda _{a},\lambda _{b})=(1,-1,-1)$%
. The numbers near the curves are the values of the total angular momentum $%
j $. The full and dashed curves correspond to the positive ($\kappa =+$) and
negative ($\kappa =-$) energy modes.

If bound states are present their contribution should be added to the
right-hand side of (\ref{jmu}). The corresponding mode functions are given by%
\begin{equation}
\psi _{\sigma }^{(\mathrm{b},\kappa )}(x)=C_{\kappa }^{(b)}e^{iq(j+\chi
)\phi -\kappa iEt}\left(
\begin{array}{c}
G_{\beta _{j},\beta _{j}}^{(a)}(\eta ,\eta r/a)e^{-iq\phi /2} \\
-\frac{e^{iq\phi /2}\eta /a}{\kappa E+sm}G_{\beta _{j},\beta _{j}+\epsilon
_{j}}^{(a)}(\eta ,\eta r/a)%
\end{array}%
\right) ,  \label{psib}
\end{equation}%
with the energy $E=\sqrt{m^{2}-\eta ^{2}/a^{2}}$. Here the function $%
G_{\beta _{j},\mu }^{(u)}(x,y)$ is defined by (\ref{Gen}) with $f_{\beta
_{j}}^{(u)}(z)$ obtained from (\ref{fnu}) by the replacement (\ref{replb}).
From the condition (\ref{Norm}) for the normalization coefficient we find%
\begin{equation}
\left\vert C_{\kappa }^{(b)}\right\vert ^{2}=\frac{q\eta ^{2}}{4\pi a^{2}E}%
\left[ B_{a}-\frac{B_{b}I_{\beta _{j}}^{(a)2}\left( \eta \right) }{I_{\beta
_{j}}^{(b)2}\left( \eta b/a\right) }\right] ^{-1},  \label{Ckb}
\end{equation}%
where $B_{u}$ is defined in accordance with (\ref{Bu}).

Substituting the mode functions into the mode sum (\ref{VEVj1}) for the
contribution of the bound states to the VEV $\langle j^{\mu }\rangle $ we get%
\begin{equation}
\langle j^{\mu }\rangle _{(\mathrm{b})}=-\frac{eq}{8\pi }\sum_{j}\sum_{%
\kappa =\pm }\,\frac{\left( \kappa sm-E\right) w_{\mu ,\beta _{j}}^{(\mathrm{%
b})}(\eta )/E}{B_{a}-B_{b}I_{\beta _{j}}^{(a)2}\left( \eta \right) /I_{\beta
_{j}}^{(b)2}\left( \eta x\right) },  \label{jb}
\end{equation}%
with $\mu =0,2$ and with the functions%
\begin{eqnarray}
w_{0,\beta _{j}}^{(\mathrm{b})}(\eta ) &=&\left( sm+\kappa E\right) G_{\beta
_{j},\beta _{j}}^{(a)2}(\eta ,\eta r/a)+\left( sm-\kappa E\right) G_{\beta
_{j},\beta _{j}+\epsilon _{j}}^{(a)2}(\eta ,\eta r/a),  \notag \\
w_{2,\beta _{j}}^{(\mathrm{b})}(\eta ) &=&-\frac{2\eta }{ar}G_{\beta
_{j},\beta _{j}}^{(a)}(\eta ,\eta r/a)G_{\beta _{j},\beta _{j}+\epsilon
_{j}}^{(a)}(\eta ,\eta r/a).  \label{wb}
\end{eqnarray}%
The total current density is the sum of the parts (\ref{jmu}) and (\ref{jb}%
). On the edges of the ring one gets simplified expression for the charge
density:%
\begin{equation}
\langle j^{0}\rangle _{(\mathrm{b}),r=u}=-\frac{eq}{4\pi a^{2}}%
\sum_{j}\sum_{\kappa =\pm }\frac{\eta ^{2}}{E}\,\frac{\kappa I_{\beta
_{j}}^{(a)2}\left( \eta \right) /I_{\beta _{j}}^{(b)2}\left( \eta u/a\right)
}{B_{a}-B_{b}I_{\beta _{j}}^{(a)2}\left( \eta \right) /I_{\beta
_{j}}^{(b)2}\left( \eta b/a\right) },  \label{j0bu}
\end{equation}%
where $u=a,b$. For the azimuthal current density on the edges we have the
relation (\ref{j2u}).

For the evaluation of the sum over $l$ in (\ref{jmu}) we again can apply the
Abel-Plana type formula (\ref{Sum}). However, in the presence of the modes $%
\gamma a=i\eta $ the summation formula (\ref{Sum}) is modified: an
additional term appears in the right-hand side coming from the poles $z=\pm
i\eta $. The derivation of the summation formula from the generalized
Abel-Plana formula is similar to that for (\ref{Sum}) presented in \cite%
{Saha08book}. The difference is that now the function $g(z)$ in the
generalized Abel-Plana formula has poles $z=\pm i\eta $ on the imaginary
axis. In the corresponding integral these poles should be avoided by small
semicircles in the right half plane with the centers at $z=i\eta $ and $%
z=-i\eta $. The contributions of the integrals over these semicircles are
combined, up to the coefficient $-\pi ^{2}/4$, as the term%
\begin{equation}
i{{\frac{I_{\nu }^{(b)}(\eta b/a)}{I_{\nu }^{(a)}(\eta )}}}\frac{{{w(i\eta )}%
-w(-i\eta )}}{\partial _{z}G_{\beta _{j}}(z,zb/a)|_{z=\eta }}.
\label{BScontr}
\end{equation}%
Now, the summation formula for the series over the positive roots $\gamma
_{l}$ is obtained from (\ref{Sum}) by adding to the right-hand side of that
formula the term (\ref{BScontr}).

After the application of the summation formula (\ref{Sum}) with the
additional term (\ref{BScontr}) in the right-hand side, the contribution to
the current density from the modes with $\gamma =\gamma _{l}$ is given by (%
\ref{jmu2}) plus the part coming from (\ref{BScontr}). By taking into
account that $w_{\mu ,\beta _{j}}(ze^{-\pi i/2})=-w_{\mu ,\beta
_{j}}(ze^{\pi i/2})$ for $z<m_{a}$, we can see that the additional term in
the VEV $\langle j^{\mu }\rangle $ is presented as
\begin{equation}
\frac{qe}{4\pi a^{2}}\sum_{j}\sum_{\kappa =\pm }\,{{\frac{I_{\nu
}^{(b)}(\eta b/a)}{I_{\nu }^{(a)}(\eta )}}}\frac{\eta w_{\mu ,\beta _{j}}^{(%
\mathrm{b})}(\eta )/E}{\partial _{z}G_{\beta _{j}}(z,zb/a)|_{z=\eta }}.
\label{BScontr2}
\end{equation}%
By using the definition (\ref{Gb}) and the fact that $z=\eta $ is the zero
of the function $G_{\beta _{j}}(z,zb/a)$, one can show that the derivative
in (\ref{BScontr2}) is given by%
\begin{equation}
\partial _{z}G_{\beta _{j}}(z,zb/a)|_{z=\eta }=\frac{2\kappa a^{-2}\eta }{%
sm-\kappa E}\left[ B_{a}\frac{I_{\beta _{j}}^{(b)}\left( \eta b/a\right) }{%
I_{\beta _{j}}^{(a)}\left( \eta \right) }-B_{b}\frac{I_{\beta
_{j}}^{(a)}\left( \eta \right) }{I_{\beta _{j}}^{(b)}\left( \eta b/a\right) }%
\right] .  \label{DerG}
\end{equation}%
Substituting this into (\ref{BScontr2}) we see that the contribution (\ref%
{BScontr2}) is the same as (\ref{jb}) but with the opposite sign. From here
we conclude that the contribution of the bound states to the total VEV $%
\langle j^{\mu }\rangle $ is cancelled by the contribution of the additional
term (\ref{BScontr}) in the summation formula for the modes $\gamma _{l}$.
Hence, all the representations for the charge and current densities given
above, started from (\ref{jmu2}), are valid in the case of the presence of
bound states as well.

\section{Special mode and its contribution to the VEVs}

\label{sec:Sp}

For half-integer values of the parameter $\alpha $ there is a special mode
corresponding to $j=-\alpha $. For this mode the upper and lower components
of the spinor are expressed in terms of the cylinder functions with the
orders $\pm 1/2$ and the mode functions obeying the boundary condition (\ref%
{BCs}) on the edge $r=a$ have the form%
\begin{equation}
\psi _{(\mathrm{s})\sigma }^{(\kappa )}(x)=C_{(\mathrm{s})}\sqrt{\frac{%
E+\kappa sm}{\phi _{0}r\left( b-a\right) E}}e^{iq(\chi -\alpha -1/2)\phi
-\kappa iEt}\left(
\begin{array}{c}
\cos \left[ \gamma \left( r-a\right) +\gamma _{a}\right] \\
\frac{\gamma e^{iq\phi }}{\kappa E+sm}\sin \left[ \gamma \left( r-a\right)
+\gamma _{a}\right]%
\end{array}%
\right) \ ,  \label{psis}
\end{equation}%
with $\gamma _{a}$ defined by the relations%
\begin{equation}
\sin \gamma _{a}=-\sqrt{\frac{E+\kappa sm}{2E}},\;\cos \gamma _{a}=\kappa
\lambda _{a}\sqrt{\frac{E-\kappa sm}{2E}}.  \label{ga}
\end{equation}%
The coefficient is given by
\begin{equation}
C_{(\mathrm{s})}=\left\{ 1+\frac{\kappa sm}{2zE}\left[ \sin \left(
2z+2\gamma _{a}\right) +\kappa \lambda _{a}\frac{\gamma }{E}\right] \right\}
^{-1/2},  \label{Cs}
\end{equation}%
with $z=\gamma (b-a)$. From the boundary condition (\ref{BCs}) at $r=b$ we
get the equation for the eigenvalues of $\gamma $:%
\begin{equation}
\kappa \left( \lambda _{a}+\lambda _{b}\right) \left( \cos z+\lambda _{a}s%
\frac{m}{\gamma }\sin z\right) +\left( 1-\lambda _{a}\lambda _{b}\right)
\frac{E}{\gamma }\sin z=0.  \label{Eigs}
\end{equation}%
The positive roots of this equation will be denoted by $\gamma =\gamma
_{l}^{(\mathrm{s})}$, $l=1,2,3,\ldots $.

On the base of the modes (\ref{psis}), for the contribution of the special
mode to the charge density one gets%
\begin{equation}
\langle j^{\mu }\rangle _{(\mathrm{s})}=-\frac{e}{2\phi _{0}\left(
b-a\right) r}\sum_{l=1}^{\infty }\sum_{\kappa =-,+}\kappa C_{(\mathrm{s}%
)}^{2}w_{(\mathrm{s})\mu }(\gamma _{l}^{(\mathrm{s})}),  \label{jmus}
\end{equation}%
where%
\begin{eqnarray}
w_{(\mathrm{s})0}(\gamma ) &=&1+\frac{sm}{E^{2}}\left\{ \lambda _{a}\gamma
\sin \left[ 2\gamma \left( r-a\right) \right] -sm\cos \left[ 2\gamma \left(
r-a\right) \right] \right\} ,  \notag \\
w_{(\mathrm{s})2}(\gamma ) &=&-\frac{\gamma }{rE^{2}}\left\{ sm\sin \left[
2\gamma \left( r-a\right) \right] +\lambda _{a}\gamma \cos \left[ 2\gamma
\left( r-a\right) \right] \right\} .  \label{ws}
\end{eqnarray}%
As it is seen from (\ref{Eigs}), for the specification of the eigenvalues $%
\gamma _{l}$ two cases should be considered separately.

In the case $\lambda _{b}=-\lambda _{a}$ the equation for $\gamma $ is
reduced to $\sin [\gamma (b-a)]=0$ with the eigenvalues $\gamma _{l}^{(%
\mathrm{s})}=\pi l/(b-a)$, $l=1,2,\ldots $. For these modes $C_{(\mathrm{s}%
)}^{2}=1$ and, hence, in (\ref{jmus}) the factor $C_{(\mathrm{s})}^{2}w_{(%
\mathrm{s})\mu }(\gamma _{l}^{(\mathrm{s})})$ is the same for the positive
and negative energy modes. Consequently, the contributions $\langle j^{\mu
}\rangle _{(\mathrm{s})}$ for both the charge and current densities are zero
because of the cancellation between the positive and negative energy modes.

For $j=-\alpha $ and $\lambda _{b}=-\lambda _{a}$, in addition to the modes
with positive $\gamma $ there is a zero energy mode with $E=0$ and $\gamma
=im$. The corresponding normalized mode function is given by%
\begin{equation}
\psi _{(\mathrm{s})}^{(0)}=\frac{\sqrt{m/(\phi _{0}r)}e^{\lambda
_{a}smr+iq\left( \chi -\alpha -1/2\right) \phi }}{\left[ \lambda _{a}s\left(
e^{2\lambda _{a}smb}-e^{2\lambda _{a}sma}\right) \right] ^{1/2}}\left(
\begin{array}{c}
1 \\
-\lambda _{a}e^{iq\phi }%
\end{array}%
\right) ,  \label{psiE0}
\end{equation}%
and $\lambda _{b}=-\lambda _{a}$. For the contribution of this mode to the
charge density we get
\begin{equation}
\langle j^{0}\rangle _{\mathrm{(s)}}^{(0)}=\pm \frac{e}{\phi _{0}r}\frac{%
\lambda _{a}sme^{2\lambda _{a}smr}}{e^{2\lambda _{a}smb}-e^{2\lambda _{a}sma}%
},  \label{j0s10}
\end{equation}%
and for the azimuthal current density one has $\langle j^{2}\rangle _{%
\mathrm{(s)}}^{(0)}=-\lambda _{a}\langle j^{0}\rangle _{\mathrm{(s)}%
}^{(0)}/r $ for all values $a\leq r\leq b$. The reason for the appearance of
two signs in the presence of the fermionic zero mode is the same as that
discussed in \cite{Jack76}

In the case $\lambda _{b}=\lambda _{a}$, the boundary condition (\ref{Eigs})
leads to the equation
\begin{equation}
\cos z+(\lambda _{a}sm/\gamma )\sin z=0,  \label{Eigs2}
\end{equation}%
with $z=\gamma (b-a)$. It is the same for the positive and negative energy
modes. Note that the equation (\ref{Eigs2}) coincides with the eigenvalue
equation for a finite length cylindrical tube (see \cite{Bell13} for the
case $\lambda _{a}s=1$). For the solutions of (\ref{Eigs2}) the expression
for (\ref{Cs}) is simplified to $C_{(\mathrm{s})}^{-2}=1-\sin \left(
2z\right) /2z$ and, again, is the same for the modes $\kappa =+$ and $\kappa
=-$. Hence, as in the previous case, the contributions of the positive and
negative energy modes cancel each other in the VEVs (\ref{jmus}).

Concluding the analysis in this section, for half-integer values of $\alpha $
the special mode with the angular momentum $j=-\alpha $ does not contribute
to the VEVs of the charge and current densities in the case $\lambda
_{b}=\lambda _{a}$. In the case $\lambda _{b}=-\lambda _{a}$ the only
nonzero contributions come from the zero energy mode (\ref{psiE0}). For the
charge density that contribution is given by (\ref{j0s10}).

\end{document}